# Tuning charge-transport properties and magnetic order in metallic $EuTiO_{3-\delta}$

Xing He[1,*,†], Chiou Yang Tan[1], Issam Khayr[1], Zach Van Fossan[2], Richard J. Spieker[1], Dayu Zhai[1], Sarah Anderson[1], Dinesh Shukla[1,3], Suchismita Sarker[4], Javier Garcia-Barriocanal[5], Turan Birol[2], and Martin Greven[1,†]

[1]School of Physics and Astronomy, University of Minnesota, Minneapolis, MN 55455, USA

[2]Department of Chemical Engineering and Materials Science, University of Minnesota, Minneapolis, MN 55455, USA

[3]UGC-DAE Consortium for Scientific Research, Indore, Madhya Pradesh 452001, India

[4]Cornell High Energy Synchrotron Source, Cornell University, Ithaca, NY 14853, USA

[5]Characterization Facility, University of Minnesota, Minneapolis, MN 55455, USA

*Present address: Department of Chemistry, University of Pennsylvania, Philadelphia, PA 19104, USA

[†]Correspondence to: xinghe25@sas.upenn.edu, greven@umn.edu

**Abstract**

The stoichiometric antiferromagnetic insulator $EuTiO_3$ is proximate to a ferroelectric phase. Whereas cation substitution has been used as a tuning parameter to introduce charge carriers and manipulate the magnetism, the effects of oxygen-vacancy doping have been less explored. Here we report a detailed study of the charge transport and magnetic properties of metallic, oxygen-vacancy-doped $EuTiO_{3-\delta}$. Using $CaH_2$ as an oxygen getter to achieve a higher carrier concentration than previously reported, we find that the phase diagram of the oxygen-vacancy-doped system is distinct from that obtained via cation doping. In particular, we uncover a change from antiferromagnetic to ferromagnetic order in the metallic state, with a maximum Curie temperature of $T_C \approx 11$ K at the highest carrier concentration of $n \approx 10^{21}$ cm$^{-3}$. These findings are supported by density functional theory calculations, which indicate a significant change in the nearest-neighbor magnetic exchange constant with increasing electron doping. We also present x-ray diffuse scattering and complementary first-principles results that reveal that, similar to the prominent incipient ferroelectric perovskite $SrTiO_3$, the data for $EuTiO_3$ are consistent with thermal diffuse scattering and with the absence of quasi-elastic contributions. Finally, we report specific heat measurements that confirm the magnetic transition temperatures deduced from magnetization measurements and corroborate the lattice dynamics picture inferred from the diffuse scattering data.

## I. Introduction

Ferroelectric (FE) materials that approach a paraelectric (PE) phase at zero temperature have garnered interest due to their proximity to a quantum phase transition [1,2]. Two prominent examples are the perovskite oxides $SrTiO_3$ (STO) and $KTaO_3$ (KTO), which exhibit a zone-center soft mode whose frequency remains nonzero at low temperatures, accompanied by a saturation of the permittivity, a phenomenon known as incipient ferroelectricity [3–7]. Notably, the emergence of bulk superconductivity in STO at low carrier concentrations [8,9] and of interfacial superconductivity in KTO, introduced via gating and at heterostructure interfaces [10,11], has been proposed to be closely related to FE soft-mode and Rashba spin-orbit coupling effects [8,12–16]. In the case of STO, alloying with Ca or Ba [17,18], $^{18}O$ isotope substitution [19,20], as well as strain [21,22] have been shown to introduce ferroelectricity and to enhance the superconducting transition temperature. Similarly, spin fluctuations and their connection to superconductivity have been extensively studied in magnetic materials [23,24]. An interesting possibility that involves both FE and magnetic fluctuations is multiferroic quantum criticality, which could potentially provide a novel platform for the exploration of the interplay of both properties and of functional behavior under applied fields [5].

Stoichiometric, insulating $EuTiO_3$ (ETO) features a band gap in the 0.2-0.9 eV range [25–28]. Optical measurements typically report a higher gap of ~0.8-0.9 eV [25,28], whereas electrical measurements and density functional theory (DFT) give smaller values of ~0.2-0.3 eV [26,27]. Like STO and KTO [6,33], ETO is an incipient FE with an incomplete zone-center transverse optic (TO) mode softening upon cooling [29–31]. Yet ETO also shows antiferromagnetic (AFM) order associated with $Eu^{2+}$ moments, with a Néel temperature of $T_N \sim 5.5$ K [32,33]. The proximity to both FE and magnetic transitions at zero temperature makes ETO a promising multiferroic quantum criticality candidate, with alloying and strain as possible tuning parameters [5]. In fact, it was recently reported that tuning through a quantum critical point was achieved via alloying, for the solid solution $Eu_{0.3}Ba_{0.1}Sr_{0.6}TiO_3$ [34]. Moreover, due to spin-phonon coupling, ETO exhibits magnetoelectric effects: the dielectric constant decreases at the Néel temperature and is enhanced in an applied magnetic field [35]. Furthermore, under biaxial strain, the ground state of ETO can transform from AFM-PE to ferromagnetic-ferroelectric (FM-FE), as shown both experimentally and via first-principles calculations [36–38], with a colossal magnetoelectric response at the AFM-PE / FM-FE boundary [39].

Metallic, electron-doped ETO exhibits itinerant electrons in Ti-3*d* conduction bands, which affects the magnetic properties, as Eu-4*f* orbitals hybridize with Ti-3*d* orbitals [40]. An increase in the carrier concentration can shift the ground-state spin configuration from AFM to FM, and thus turn the system into a FM metal, as indicated theoretically [41] and demonstrated experimentally, with electron doping achieved via substitution on both the perovskite ($ABO_3$) A- (La [42–44], Gd [45]) and B-site (Nb [46], Al [47]), and via hydrogen substitution ($EuTiO_{3-x}H_x$ [48]). However, whereas FM order has been observed in oxygen-deficient thin films [51,52], this has not yet been explored in bulk crystals [53]. Another intriguing aspect of doped ETO is the evidence of incommensurate superstructure correlations due to antiferrodistortive and competing instabilities, which were proposed to be highly sensitive to defects such as $Eu^{3+}$ and oxygen vacancies [54,55], although the exact nature of these correlations has remained unclear. A comprehensive investigation of the transport, magnetic, and structural properties of bulk oxygen-vacancy doped (OVD) ETO is therefore highly desirable.

Here we investigate the charge transport and thermodynamic properties of OVD $EuTiO_{3-\delta}$ single crystals with increasing oxygen deficiency. In addition, we present synchrotron x-ray diffuse scattering data for stoichiometric, insulating ETO, and we complement our experimental work with first-principles (DFT + $U$) calculations. The diffuse scattering data confirm the high structural quality of our samples and, together with the DFT + $U$ results, indicate that the structural response originates from phonon contributions, with a lattice dynamics that is similar to STO, as both perovskites exhibit a cubic-tetragonal phase transition associated with a zone-boundary soft-mode [40,56]. We find that oxygen-vacancy doping with calcium hydride ($CaH_2$) as a getter material can result in electron-carrier concentrations that exceed previously reported values [53] and that the magnetic ground state evolves from AFM to FM. This change in the metallic state is supported by our DFT + $U$ calculation, which indicates a significant modification of the Eu-Eu nearest-neighbor exchange constant with increasing doping. The theoretical analysis furthermore indicates that, although the three-fold degeneracy of the conduction band at the Γ-point is lifted by spin-orbit coupling (SOC), nearly all doped samples studied in the present work fall into the regime in which all three bands are active. Finally, specific heat measurements corroborate the lattice dynamics picture inferred from diffuse scattering and confirm the magnetic transition temperatures observed via magnetometry. The present work lays the foundation for further studies. For example, analogous to Ba/Ca and Nb-doped STO [17,18], Ba/Ca-substituted $EuTiO_{3-\delta}$ might host a polar metallic state. In this state, spin-phonon coupling might allow the manipulation of polar properties with an applied magnetic field [35], while itinerant electron-spin coupling could enable magnetic-field-tuning of charge-transport properties [43,45,46]. Moreover, our diffuse x-ray scattering data provide a benchmark for future studies of antiferrodistortive and competing instabilities and their relationship with $Eu^{3+}$ defects and oxygen vacancies.

This paper is organized as follows: In Section II, we briefly describe the techniques used for sample preparation and measurement, as well as DFT + $U$ calculations, and we provide additional details in the Supplemental Material [57]. Section III describes our results in five sub-sections: structural characterization and lattice dynamics, charge transport and insulator-metal transition, magnetic phase diagram, specific heat, and carrier concentration effects on spin interaction and ground state. We then discuss and summarize our results in Sections III and IV.

## II. Methods

Single crystals of ETO were grown with the floating-zone method. The crystals were cut into small samples, and their orientation was determined via Laue x-ray diffraction. Other techniques, including in-house powder and single-crystal x-ray diffraction as well as synchrotron x-ray diffuse scattering were used to characterize the samples, as described below. We used $CaH_2$ as a getter material to create oxygen vacancies in ETO. Nominally stoichiometric, as-grown ETO was sealed with $CaH_2$ powder (without direct contact) in quartz tubes under vacuum (pressure below 100 mTorr), with a 1:12 mass ratio of ETO to $CaH_2$. The sealed quartz tubes were heated for 12 h at temperatures between 700°C and 1050°C, depending on the desired carrier concentration, with higher temperatures resulting in higher oxygen vacancy densities. Resistivity and Hall measurements were conducted with the standard four-point van der Pauw method using a Quantum Design, Inc. Physical Property Measurement System (PPMS) in AC drive mode. A wire bonder was used to make contacts with aluminum wires. For metallic samples, the contact resistance was about 1 Ω. DC magnetic-susceptibility measurements were performed with a Quantum Design, Inc. Magnetic Property Measurement System (MPMS). A 20 Oe DC field was applied along either <100> or <110> directions. Synchrotron x-ray diffuse scattering measurements were performed at beamline QM2 ID4B of the Cornell High Energy Synchrotron Source (CHESS). DFT + $U$ calculations were performed at the Agate cluster of the Minnesota Supercomputing Institute. Electronic and magnetic properties were calculated using VASP [58,59] in the cubic phase of ETO. The phonon properties were calculated with the finite displacement method, on the tetragonal phase with a 2×2×2 supercell, using Quantum Espresso [60,61] and the Phonopy [62] package. Specific heat measurements were made in a PPMS using relaxation calorimetry. Apiezon N grease was used to affix crystals to PPMS platforms for measurements between 1.85 K and 300 K, and 2% temperature pulses were used. At each temperature, three measurements were performed in order to achieve improved accuracy. Other experimental and computational details are described in the Supplemental Material [57].

## III. Results

### A. Structural characterization and lattice dynamics

The perovskite ETO undergoes a phase transition from simple cubic ($Pm\overline{3}m$ structure) to tetragonal (*I4/mcm* structure) somewhat below room temperature [31,55,56,63,64] (Fig. 1(a)). The conventional tetragonal cell parameters are related to the cubic lattice constant $a_0$, with $a \sim \sqrt{2}a_0, b \sim \sqrt{2}a_0,$ and $c \sim 2a_0$. This phase transition involves octahedral rotation, as in the case of STO [65,66], and it is associated with the condensation of a zone-boundary $R$-point mode [31]. Furthermore, undoped ETO undergoes a PM-AFM transition with a Néel temperature of $T_N \sim 5.5$ K. The observed $G$-type AFM order is illustrated in Fig. 1(b), with spins aligned along $<100>_{cubic}$ (equivalently, $<110>_{tetragonal}$), as determined from neutron powder diffraction and resonant x-ray scattering [32]. In this $G$-type AFM state, nearest-neighbor and second-nearest-neighbor Eu moments are antiparallel and parallel to each other, respectively. The three nearest exchange interactions $J_1$, $J_2$, and $J_3$ are indicated in Fig. 1(b).

Our as-grown single crystals were characterized with several x-ray techniques. The Laue backscattering x-ray pattern of an ETO crystal is shown in Fig. 1(c). The resultant diffraction pattern exhibits sharp Bragg peaks. Additionally, Cu K-α powder x-ray diffraction (XRD) was performed on ground single-crystal samples, and the pattern shown in Fig. 1(d) matches well with the calculated $2\theta$ positions. The high crystalline quality was further confirmed via single-crystal XRD characterization (see Fig. S1 [57]) and synchrotron x-ray measurements (see below).

In order to characterize the atomic dynamics and probe structural correlations, we obtained synchrotron x-ray diffuse scattering data for an as-grown ETO crystal (additional data for an OVD ETO crystal are shown in Fig. S2 [57]). While x-ray diffuse scattering was measured previously [54], this prior work did not cover a sufficiently large momentum space volume to enable a detailed characterization of the lattice dynamics. Figure 2(a) compares the diffuse scattering pattern in the (*HK*0) plane, obtained at 175 K and 300 K, with a DFT + *U* calculation of the thermal diffuse scattering (TDS) due to phonons. Since the cubic phase of ETO exhibits an unstable phonon mode [67], it is necessary to calculate phonons in the tetragonal phase for TDS simulations. The resultant phonon dispersions and phonon density of states (PDOS) are shown in Fig. 2(b). We note that, in Fig. 2(b), we use high-symmetry *k*-point labels for the body-centered tetragonal cell ($c/a > 1$), whereas in the remainder of the manuscript the wavevectors are defined with respect to the five-atom perovskite cell. Phonon eigenvalues and eigenvectors were subsequently used to calculate the dynamic structure factor, $S(\mathbf{Q},\omega)$, where $\mathbf{Q}$ and $\hbar\omega$ are the momentum and energy transfers, respectively, as detailed in [57]. The TDS simulations were then generated by integrating the calculated $S(\mathbf{Q},\omega)$ over the $\hbar\omega$ = 0 - 30 meV energy range. In the (*HK*0) plane, streaks between Bragg peaks along $\langle 110\rangle_{\text{cubic}}$ are clearly seen in both the diffuse scattering data and the results of the theoretical model. The observed diffuse scattering pattern is similar to that for STO [68,69], owing to the same perovskite structure and structural instabilities associated with octahedral rotations, indicative of closely similar atomic dynamics. By investigating the phonon eigenvector components along high-symmetry directions, as shown in Fig. 2(b) (see also Fig. S8 [57]), it can be deduced that the diffuse scattering along $\langle 110\rangle_{\text{cubic}}$ ($\Gamma$ - $M$) originates from nearly dispersionless modes around 11 meV that are dominated by Eu motion. These low-energy modes correspond to the first peak in the atom-projected phonon density of states (DOS) in Fig. 2(b), in agreement with a Eu-specific PDOS measurement [64], and consistent with our DOS calculation and thermal diffuse scattering result. As shown below, our specific heat data provide additional thermodynamic validation of this low-energy phonon contribution.

Figure 2(c,d) compares the experimental diffuse scattering patterns for the half-integer planes (*HK*2.5) and (*HK*0.5) at 175 K and 300 K with our DFT + *U* result. We note that the latter does not capture the experimental background intensity modulation at low $\mathbf{Q}$ that results from strong air scattering. These data again show similarities with STO [68,69] and reveal several characteristic features. First, the diffuse peaks at the half-integer positions ($h/2$ $k/2$ $l/2$) (with integer $h$, $k$, and $l$) are directly related to the cubic-tetragonal transition associated with the soft acoustic phonon at *R*-points. At 300 K, this low-energy phonon has not yet condensed, leading to thermal diffuse scattering. In contrast, at 175 K, below $T_s$, the phonon has condensed, and we observe superlattice Bragg peaks at the *R* points (see also Fig. 2(e)). The eigenvector of the *R*-point phonon corresponds to out-of-phase rotation of the octahedra in the *a*-*b* plane, as depicted in Fig. 2(f). Neighboring octahedra rotate in opposite directions about the *c*-axis, in a *G*-type pattern. This distortion drives the transition from the cubic $Pm\overline{3}m$ to tetragonal *I4/mcm* structure. Second, we observe weak cross-shaped diffuse scattering near the *R*-points along <100>$_{cubic}$ (Fig. 2(c)), e.g., around **Q** = (-1.5 ±1.5 2.5). The cross-shaped features can be better seen in Fig. 3(a-b), and we attribute them to phonons with wavevectors between the *M* and *R* points. These phonons also correspond to octahedral rotations, but with a different spatial periodicity than simple in-phase or out-of-phase rotations (i.e., with long-wavelength modulations along the axis of rotation). These features are also present in the calculated oxygen-only diffuse scattering map in Fig. 3(c). Third, streaks through ($h$ $k$ $l$/2) positions (with integer $h$, $k$, and $l$) and the curvilinear features of the diffuse pattern in half-integer planes (Fig. 2 (c) and (d)), analogous to STO [68], are well reproduced by theoretical modeling and attributed to Eu and Ti distortions (Fig. 3(d) and Fig. S9 [57]). Fourth, we did not observe incommensurate peaks in the as-grown ETO sample (at 175 K and 300 K), likely as a result of differences in defect types or densities compared to prior work [54,55]. Lastly, we note that we also measured a metallic OVD ETO sample and observed closely similar diffuse scattering patterns (Fig. S2 [57]).

**B. Charge transport and insulator-metal transition**

We conducted resistivity and Hall-effect measurements of both as-grown and OVD ETO samples with varying doping levels. The respective results are shown in Figs. 4(a) and (b). The resistivity of the as-grown sample is $\rho \approx 100\ \Omega$cm at 300 K and becomes immeasurably large at low temperatures. With increasing electron density, oxygen-deficient ETO is seen to evolve to a metallic state. At the highest doping level (Hall carrier concentration $n_H \approx 2.2 \times 10^{21}$ cm$^{-3}$), $\rho \approx 3.33 \times 10^{-4}\ \Omega$cm at 300 K, and the residual resistance ratio (RRR) is 3.3. The resistivities of the six samples with the highest doping levels exhibit a $T^2$ relationship at low temperatures and cusps between 5.5 and 12 K, indicative of a magnetic phase transition (Fig. 4(a)). The sample with the seventh-highest doping level ($n_H \approx 7.3 \times 10^{18}$ cm$^{-3}$) shows an upturn in resistivity upon cooling, deviating from typical metallic behavior. We therefore identify the metal-insulator transition (MIT) carrier concentration is between the seventh- and sixth-highest doping level ($7.3 \times 10^{18} - 5.5 \times 10^{19}$ cm$^{-3}$). For the undoped sample ($n_H < 1.7 \times 10^{16}$ cm$^{-3}$), the resistivity increases monotonically upon cooling, consistent with insulating behavior. As seen from Fig. 4(b), the as-grown sample shows a decrease in $n_H$ upon cooling, reflecting its insulating behavior, whereas the OVD samples exhibit nearly temperature-independent carrier concentrations, consistent with prior work [53].

The resistivity of the six metallic samples exhibits quadratic temperature dependence over an extended temperature range, and the results of fits to the form $\rho = \rho_0 + AT^2$ for the six highest doping levels are shown in Figs. 4(c) and (d). With increasing carrier concentration, the quadratic relationship extends to higher temperatures (~ 150 K, 220 K, 240 K, 280 K, 300 K, and 300 K; note that our data do not extend above 300 K; for additional information regarding these fits, see Figs. S17 and S18 [57]). As seen from Fig. 4(d), the extracted A($n$) values are in overall good agreement with prior results [53] and extend this earlier work to both lower and higher doping levels.

The band structure of cubic ETO with AFM order, from which parameters were obtained for the three-band model, is shown in Fig. 7(c-d). Results from DFT + $U$ calculations performed with and without SOC are given to illustrate the dependence of the Ti-$d$ band splitting on relativistic effects in the system. Interestingly, with SOC included, the Ti-$d$ bands at Γ split into a single band that is 31 meV higher than the other two lower energy bands, while the splitting of the two lowest energy bands is negligible. This behavior of cubic ETO with AFM order, including SOC is similar to that of cubic STO [70]. In STO, the tetragonality splits the two lowest energy bands further by about 5 meV [71], and we adopt this for our ETO three-band model instead of calculating tetragonal AFM ETO, for simplicity. In contrast, the band structure results obtained for ETO with FM order are different from AFM, with large splitting between all three bands (as shown in Figs. S8 and S9). See the Supplemental Material [57] for additional information and a discussion of the effect of varying the on-site Coulomb interaction on the hybridization between the Eu-4$f$ and Ti-3$d$ orbitals.

We adopted the simple parabolic three-band model of Ref. [53], but rather than using band minima parameters and effective mass values for STO [53], we adjusted these to match ETO based on our DFT calculation. The carrier concentrations of each band is:

$$n_i(E_F) = 1/3\pi \left(\frac{2m_i}{\hbar^2}\right)^{3/2} \int_{E_i}^{E_F} \sqrt{E - E_i}\, dE \tag{1}$$

with Fermi energy $E_F$, band minima at $E_0$, $E_1$, and $E_2$, effective masses $m_0$, $m_1$, and $m_2$, and total electron density $n(E_F) = n_0 + n_1 + n_2$. We used $m_0 = 1.24m_e$, $m_1 = 3.47m_e$, $m_2 = 1.62m_e$, $E_0 = 0$ meV, $E_1 = 5$ meV (approximate), and $E_2 = 31$ meV from our DFT + $U$ calculation (see also below and Figs. S13 and S14 [57]). Upon inverting $E_F(n)$, we obtain $A(n) \propto [E_F(n)]^{-2}$ and, consequently, $A \propto n^{\alpha}$, where $\alpha$ changes as $n$ increases. In the three-band regime ($E_F > E_2$), $\alpha$ initially reaches a peak value of -1 and then gradually decreases to -4/3 as $n$ increases (see also Fig. S14 [57]). Thus, we first fit the data to $A(n) = F_1 \cdot n^{-1}$ at the six highest doping levels, and then adjust the pre-factor $F_2$ of the three-band result $A(n) = F_2 \cdot n^{\alpha}$ to coincide with $F_1 \cdot n^{-1}$ at $n_H = 5.5\times 10^{19}$ cm$^{-3}$, as shown in Fig. 4(d). We note that the simple heuristic behavior $A(n) \propto n^{-1}$ is roughly in agreement with our data, but that the three-band model provides a better description as $A$ decreases faster at higher $n_H$ as $\alpha$ becomes smaller than -1 (Fig. 4(d)).

### C. Magnetic phase diagram

We conducted detailed magnetization measurements to investigate the effect of doping on the magnetic properties of OVD ETO. The field-cooled (FC) magnetic susceptibility ($\chi$), inverse susceptibility ($1/\chi$), and temperature derivative ($d\chi/dT$) are presented in Figs. 5(a-c), respectively. Below $n_H = 3.5 \times 10^{20}$ cm$^{-3}$, $\chi(T)$ exhibits a maximum indicative of AFM order [26,72,73]. The Néel temperature extracted in this manner is nearly independent of carrier concentration and in the $T_N = 5.2$ - $5.5$ K range, consistent with prior studies of La-doped [42,43,50], Al-doped [47], Nb-doped [74], and OVD [53] ETO that found mild or no doping dependences. For $n_H > 3.5 \times 10^{20}$ cm$^{-3}$, the data are consistent with FM rather than AFM order, with a nearly temperature-independent susceptibility below $T_C$ (Fig. 5(a)). Similar to work on FM $EuTi_{1-x}Al_xO_3$ [74] and $EuTiO_{3-x}H_x$ [48], we extract the Curie temperature from the minimum of $d\chi/dT$ (Fig. 5(c)).

The susceptibility and its temperature derivative for the sample with the intermediate carrier concentration $n_H = 3.5 \times 10^{20}$ cm$^{-3}$ are shown separately in Fig. 5(d). $\chi(T)$ still exhibits a kink at ~ 5.25 K, suggesting the presence of residual AFM order. In addition, $d\chi/dT$ also shows a local minimum at ~ 5.5 K, indicative of FM order. Moreover, upon further cooling, $\chi(T)$ increases monotonically, and $d\chi/dT$ remains negative and does not cross zero, indicative of paramagnetic behavior down to the lowest measured temperature of 3.8 K. We considered whether this sample resides in a miscibility gap, i.e., if its response indicates a first-order phase transition with macroscopic AFM and FM regions. To this end, we modeled this sample's susceptibility as a linear superposition of the susceptibilities of the two samples with adjacent doping levels, $n_H = 5.5 \times 10^{19}$ cm$^{-3}$ and $9.0 \times 10^{20}$ cm$^{-3}$, with respective weights of about 65% and 35% (Fig. S6 [57]). Although this superposition captures the response above ~ 5.5 K, it does not describe the observed low-temperature behavior that is indicative of a third, paramagnetic contribution. Adding an extra paramagnetic component captures this low-$T$ trend better (Fig. S6 [57]). Assuming that we can rule out macroscopic chemical inhomogeneity, the evidence for AFM, FM, and paramagnetic contributions might be understood within a three-dimensional percolation picture, whereby two distinct species (with AFM and FM order) can simultaneously percolate at intermediate concentrations and co-exist with non-percolating zero-dimensional clusters (paramagnetic regions) [75,76] . As demonstrated in Fig. S3 [57], we consistently observed this complex magnetic behavior in multiple samples with approximately the same doping level.

The susceptibility was fit to a Curie-Weiss form (Fig. 5(b) inset), and the doping dependence of the extracted Curie temperature, $\theta_{CW}$, is summarized in Fig. 5(f). Similar to our analysis of the resistivity data, we investigated how the fit temperature range affects $\theta_{CW}$. Specifically, by varying $T_{min}$ from 10 K to 20 K, and fitting the data between $T_{min}$ and 50 K, we obtained reliable error estimates (see also Fig. S5 [57]). For as-grown ETO, we find $\theta_{CW} = 2.97(4)$ K, which is overall consistent with the prior results $\theta_{CW} = 3.6$ K [28], 3.8 K [33], 3.5 K [73], and 3 K [47]. Notably, although ETO exhibits AFM order, $\theta_{CW}$ has a small positive value, likely due to magnetic frustration [47,73]. FM samples exhibit a concomitant increase of $T_C$ and $\theta_{CW}$, with $T_C = 11.4(9)$ K and $\theta_{CW} = 8.64(3)$ K at the highest electron concentration of the present study. $T_C$ exhibits an approximately linear doping dependence (see Fig. 5(e), inset). This behavior is distinct from the effects of doping via cation substitution (A-site: La [42,44,49], Gd [45]; and B-site: Nb [46], Al [47]) and hydrogen substitution ($EuTiO_{3-x}H_x$ [48])), in which cases $T_C$ was found to remain constant or decrease as $n$ increases. Figure 5(e) shows the magnetic phase diagram of OVD ETO established in the present work.

We note here that we also obtained $M$ *vs.* $H$ data at 4.2 K for the as-grown (AFM) and most highly doped (FM) samples (see Fig. S15 [57]). The observed steep increase in $M$ at low $H$ and the clear saturation at high $H$ exhibited by the $n_H = 2.2 \times 10^{21}$ cm$^{-3}$ sample indicates robust FM order. The as-grown sample exhibits a smaller slope at low fields and, moreover, a convex dependence on $H$.

The latter is consistent with the AFM ground state and associated with spin-flips, as observed previously [26,48]. For both samples, the magnetization gradually saturates at high fields and approaches the ideal $Eu^{2+}$ ($S = 7/2$) moment of 7 $\mu_B$ at 5 T. We note that the AFM sample shows a slower, more gradual change compared with the FM sample, similar to H-doped ETO [48] and Nb-doped ETO [74]. Consistent with prior studies of undoped and doped $EuTiO_3$ [26,48,77,78], we observe hardly any hysteresis. While a lack of hysteresis is expected in the AFM case, the weak hysteresis in the FM case is unusual. However, this observation for OVD ETO is consistent with prior measurements of La-doped [42,45], Nb-doped [74,77], and H-doped [48] samples. Small hysteresis loops have been observed in other Eu-base ferromagnets (e.g., $Eu_8Ga_{16}Ge_{30}$ [79]), as they are soft ferromagnets with a small coercive field.

**D. Specific heat**

We further employed specific heat measurements to probe the thermodynamic properties of undoped and OVD ETO. We measured three samples on which we had carried out the charge transport and magnetization measurements described above: the as-grown AFM crystal and the two FM samples with Hall carrier concentrations of $n_H = 2.0 \times 10^{21}$ cm$^{-3}$ and $n_H = 2.2 \times 10^{21}$ cm$^{-3}$. Figures 6(a,b) show $C_p$ and $C_p/T$, respectively. For all three samples, $C_P$ decreases upon cooling and exhibits distinct peaks indicative of their respective magnetic transition temperatures, as highlighted in the inset of Fig. 6(a). Above the magnetic transition temperatures, $C_P$ and $C_P/T$ overlap for all three samples due to the small electronic and magnetic contributions at higher temperatures. In the 12 - 300 K range, $C_P(T)$ for the undoped sample was fit to a Debye-Einstein model (Eq. S1 [57]), which captures both the dispersive acoustic and the relatively dispersionless optic phonon modes (Fig. 6(c)). The resultant characteristic temperatures are $\Theta_D = 245(10)$ K and $\Theta_E = 635(18)$ K, with corresponding energies $\hbar\omega_D = k_B\Theta_D \sim 21$ meV and $\hbar\omega_E = k_B\Theta_E \sim 55$ meV. The Debye model effectively captures the low-energy Eu and Ti modes, whereas the Einstein model captures the high-energy oxygen modes (see also Fig. 2(b)). Also shown in Fig. 6(c) is $C_p$ obtained from our DFT + $U$ phonon calculation by summing over all modes across the Brillouin zone (Eq. S2 [57]). Both the Debye-Einstein model and the DFT + $U$ calculation show good agreement with our specific heat data.

Our data for as-grown ETO are consistent with previous reports [26,64]. For FM OVD ETO, they furthermore exhibit characteristic magnetic transition temperatures consistent with our transport and magnetization results. In an effort to further explore the thermodynamics of as-grown and doped ETO and the relationship with the lattice dynamics, we show $C_p/T^3$ *vs*. $T$ in Fig. 6(d). All three samples exhibit a distinct increase at low temperatures, with respective magnetic contributions, but share a common hump at about 25 K. This hump corresponds to the dispersionless 11 meV Eu phonon mode, as will be discussed below.

**E. Carrier concentration effects on spin interaction and ground state**

In order to better understand the changes in the magnetic ground state with increasing carrier concentration, we performed DFT + $U$ calculations of ETO in the cubic phase, with $n$ varied from zero to 0.1 electrons per Eu, and calculated the first-, second-, and third-nearest-neighbor Eu-Eu exchange interactions, $J_1$, $J_2$, and $J_3$ (see Fig. 1(b)). As in previous work on cubic ETO, the on-site Coulomb interaction, $U_{Eu}$, and on-site exchange (Hund's exchange), $J_{Eu}$, were chosen to be 5.7 eV and 1.0 eV, respectively, to obtain the experimental ratio of Néel to Curie temperatures [36,39,67]. We note that $U_{Eu}$ values in the 5.57 to 8.0 eV range were used in prior first-principles studies to better capture experimental findings for the lower-symmetry tetragonal phase [22,37,41,43,80]. We found the consideration of the cubic phase to be sufficient for capturing the carrier-concentration dependence of the exchange interactions. For $J_{Eu}$, a value of 1.0 eV is a common choice [39,43,67,80]. The dependence of the exchange constants on doping is summarized in Fig. 7(a). Notably, most prior studies of undoped ETO only included exchange interactions up to the second-nearest neighbor, and, consequently, $J_1$ and $J_2$ were assigned to be negative (FM) and positive (AFM), respectively [41,81]. We emphasize that the third-nearest-neighbor exchange interaction is crucial to stabilize the AFM ground state in the undoped case [40,82]. As a result, our calculation shows that $J_1$ is positive (AFM), $J_2$ is negative (FM), and $J_3$ is positive (AFM) in undoped ETO, with $J_1$ close to zero, while the former calculation without including $J_3$ results in a negative $J_1$ [41]. As seen in Fig. 7(a), as $n$ increases, $J_1$ decreases significantly and changes sign, whereas $J_2$ and $J_3$ increase slowly. Figure 7(b) shows the energy difference between the AFM and FM states as the carrier concentration increases. In the undoped case, this difference is negative, and the AFM ground state is favored. As $n$ increases, the energy difference between the two states decreases in magnitude, changes sign at $n \sim 5.5\times 10^{20}$ cm$^{-3}$, and the FM state becomes energetically preferred. This trend agrees well with our experimental findings and with a prior computational study [41]. The calculated "critical" carrier concentration is very close to our experimental result of $n \sim 3.5 \times 10^{20}$ cm$^{-3}$. It is interesting that $J_1$ is the most sensitive to $n$, whereas $J_3$ is somewhat unaffected. This is likely a result of the geometry of the perovskite structure, which dictates that $J_3$ is dominated by Ti-mediated exchange processes (Fig. 1(a,b)), whereas $J_1$ has contributions from both O- and Ti-mediated hopping processes. The sensitivity of $J_1$ to electron-doping is equivalent to increased occupation of Ti-*d* orbitals, as discussed below.

## IV. Discussion

Seeking to relate our exchange interaction results and charge transport measurements to the electronic structure, we calculated the electronic band structure for cubic ETO with AFM order, both with and without spin-orbit coupling (SOC) effects (Figs. 7(c,d)). We performed the band structure calculations for the cubic phase, consistent with our DFT + $U$ calculations. Consideration of the higher-symmetry cubic phase reduces the number of exchange interactions that need to be accounted for, yet still provides insight into which exchange constants and mechanisms are affected by changes in the carrier concentration. In the AFM state, the spin-up and spin-down bands are degenerate, as expected. Without SOC, the lowest conduction bands are triply degenerate at the Γ point (Fig. 8(c)), and SOC lifts this degeneracy (Fig. 8(d)). The effective masses $m_0$ = $1.24m_e$, $m_1$ = $3.47m_e$, $m_2$ = $1.62m_e$, and band splitting energies $E_0$ = 0 meV, $E_1$ = 5 meV (approximate as described earlier), and $E_2$ = 31 meV were used for our simple parabolic three-band model to calculate the transport coefficient A($n$) in Fig. 4(d). The full A vs. $n$ relationship of the three-band model is illustrated in Fig. S14 [57]. The Lifshitz transition carrier concentrations $n_{c1}$ and $n_{c2}$, where the Fermi energy crosses the second and the third bands, are indicated in Figs. 5(e,f). As seen in Fig. 4(d), for metallic ETO, the observed A vs. $n$ relationship follows expectations for a three-band model, similar to metallic STO at high carrier concentrations ($n_H \geqslant 5.5 \times 10^{19}$ $cm^{-3}$) [53]. Regardless of the overall similarities (perovskite structure, incipient ferroelectricity, structural phase transition, and band structure), ETO and STO show somewhat distinct charge transport behavior. Notably, the carrier concentration in $n$-type STO can be easily tuned to the highly dilute one-band regime [9,83,84]. The sixth highest doped ETO sample lies at border between the two- and three-band regimes, whereas the five highest doped samples lie in the three-band regime. We have not succeeded in preparing metallic samples with lower carrier concentrations. We note that in our DFT + $U$ band structure calculation, the electrons are directly added to the system without oxygen vacancy defects. As the carrier concentration increases, the Fermi energy gradually crosses the first, the second, and the third conduction bands. However, OVD and cation doping introduce point disorder, and the carriers are localized at dilute doping levels, which cannot be captured solely by band physics. When the carrier concentration $n$ reaches the Mott criterion $n_c^{1/3}a_B^* \approx 0.25$, where $a_B^*$ is the effective Bohr radius, the doped electrons delocalize and become conductive [85]. ETO exhibits a considerably smaller low-temperature permittivity than STO ($\varepsilon$ ~400 vs. ~20,000 [35]) and, consequently, a smaller effective Bohr radius $a_B^* = a_0\, \varepsilon\, m_e/m_i$, where $a_0$ is the standard Bohr radius. As a result, the insulator-metal transition in OVD ETO occurs at a higher carrier concentration (between the sixth- and seventh-highest doping levels: $7.3 \times 10^{18}$ $cm^{-3}$ ~ $5.5 \times 10^{19}$ $cm^{-3}$ ) than in $n$-type STO [43,53], which has been reported to exhibit metallic behavior down to $n_H = 8.5 \times 10^{15}$ $cm^{-3}$ [86]. In perovskite oxides, where there is no sharp MIT, the carrier concentration required for metallicity is much higher than the Mott criterion [86]. In practice, $n_c^{1/3}a_B^* \approx 10$ has been able to give a good prediction for STO, KTO, and ETO [53]. When using $\varepsilon \sim 400$ and $m_1 = 1.24\, m_e$, we obtain a MIT carrier concentration $2.0 \times 10^{20}$ $cm^{-3}$, which is close to our experimental findings ($7.3 \times 10^{18}$ $cm^{-3}$ ~ $5.5 \times 10^{19}$ $cm^{-3}$). The behavior that MIT carrier concentration is much higher than the Mott criterion in STO semiconductor systems has been explained by strong compensation effects coming from the background charged defects [87]

As demonstrated by our charge transport and magnetization measurements, the magnetic order involving Eu moments exhibits a significant doping dependence. In order to further establish how the doped electrons couple with the magnetic properties, we calculated the atomic-orbital-projected band structure. As seen from Fig. 8, the low-energy conduction bands have Ti-3$d$ character, and the Eu-4$f$ band lies just below, as previously noted based on first-principles calculations [41,43]. As a result, when $n$ and the Fermi energy increase, itinerant electrons fill the Ti-3$d$ bands and interact with Eu spins, analogous to the Ruderman-Kittel-Kasuya-Yosida (RKKY) interaction, as proposed for rare-earth or Nb-doped bulk ETO and Nb-doped oxygen-deficient ETO films [45,51,88]. The nearest-neighbor interaction constant $J_1$ has contributions from both O-and Ti-mediated hopping processes, and, consequently, the itinerant electrons in the Ti-3$d$ bands cause the magnitude and sign of $J_1$ to change. We note that charge conduction, originating from Ti-3$d$ electrons, can be tuned by an external magnetic field due to the strong coupling to local spins, similar to La-doped and Nb-doped ETO [45,88]. As shown in Fig. S16 [57], we measured the longitudinal resistance $R_{xx}$ of the $n_H$ = 2.0 x $10^{21}$ $cm^{-3}$ sample in different applied magnetic fields. In zero field, $R_{xx}$ shows a slight increase upon cooling from 30 K to $T_C$ = 7.5 K, and a sharp cusp at $T_C$, followed by a decrease below $T_C$. With increasing magnetic field, the resistance close to the transition temperature is suppressed, and the cusp gradually disappears, indicating a large negative magnetoresistance. This behavior agrees well with prior studies of La-doped [42,44,45] and H-doped [48] ETO, and it reflects the FM nature of OVD ETO at higher electron concentrations as well as the coupling between itinerant electrons in Ti-3$d$ orbitals and Eu moments.

Lastly, we analyzed the low-temperature properties of the specific heat and the corresponding lattice dynamics. In Fig. 6(d), $C_P/T^3$ *vs.* $T$ of one undoped and two metallic OVD ETO are compared with a DFT + $U$ phonon calculation of undoped ETO. As noted, $C_P/T^3$ for the three measured samples shows somewhat distinct increases below ~ 12 K due to magnetic contributions, but the data share a common hump near 25 K. To elucidate the origin of this hump, we first fit $C_P$ in the 12 - 100 K range to a Debye-only model. This model, which captures the contribution from acoustic modes, approaches a constant in the low-temperature limit and fails to reproduce the hump at ~ 25 K. To account for this feature, we include an additional Einstein term with $\Theta_E$ ~ 128 K (following a similar approach used for undoped/doped STO [89]). This Einstein temperature corresponds to a flat phonon mode with energy $\hbar\omega_E = k_B\Theta_E$ = 11 meV, and successfully reproduces the observed behavior of $C_P/T^3$. This low-energy Einstein mode is distinct from the higher Einstein temperature $\Theta_E$ = 635(18) K obtained from full-temperature fits, which accounts for the high-energy oxygen vibrations. Instead, the 11 meV mode corresponds to the Eu-dominated peak in the phonon density of states as seen in Fig. 2(b) and is independent of doping, consistent with its nonpolar character and its distinction from the ferroelectric soft mode. We further compare this behavior with our DFT + $U$ phonon calculation. While the calculations contain a magnetic specific heat contribution, and therefore do not capture the low-temperature upturn associated with magnetic order, they reproduce both the Debye-like low-temperature flattening and the hump near 25 K that arises from the flat phonon branch at ~11 meV. This agreement supports our assignment of the 25 K feature to the low-energy Eu-dominated phonon mode.

The hump observed in $C_P/T^3$ of ETO is consistent with prior measurements on undoped ETO [26]. Similarly, undoped and doped STO [89] also exhibit a hump, which is independent of doping and corresponds to the Sr-dominated phonon peak at ~12.9 meV in the phonon density of states [90]. In ETO, this feature occurs at slightly lower energy, reflecting the larger atomic mass of Eu relative to Sr, and likewise exhibits no doping dependence. Our analysis of the low-temperature specific heat, therefore, highlights the close similarity of the lattice dynamics in ETO and STO. Moreover, the quantitative agreement between DFT + $U$ calculation, including the reproduction of the Einstein-like feature near 25 K, provides further validation of the computed phonon density of states and the interpretation of the synchrotron x-ray diffuse scattering results.

## V. Conclusion

In conclusion, we successfully employed $CaH_2$ as an oxygen getter to tune the carrier concentration of the perovskite $EuTiO_{3-\delta}$ up to an unprecedented electron density of $n \approx 10^{21}$ cm$^{-3}$. This allowed us to uncover a magnetic ground-state transition of the metallic system from AFM to FM order at $n_H \sim 3.5 \times 10^{20}$ cm$^{-3}$, with a maximum Curie temperature of $T_C \sim 11$ K at the highest electron density, comparable to prior results obtained via cation substitution. A magnetic ground-state change at about the same doping level is also predicted by our complementary DFT + $U$ calculation. Our theoretical analysis of the charge transport properties, considering spin-orbit coupling, indicates three non-degenerate active bands in the doping range of the present study and yields an $A$-coefficient that is consistent with the data. We find good agreement between our x-ray diffuse scattering data and thermal diffuse scattering calculation, which indicates that the response is dominated by phonon contributions. The diffuse scattering patterns share similarities with $SrTiO_3$, which reflects the closely similar lattice dynamics of these two perovskites, with a common cubic-tetragonal phase transition associated with a zone-boundary soft-phonon mode. Moreover, our diffuse scattering and DFT + $U$ phonon calculation reveals a dispersionless Eu mode at ~11 meV, consistent with our specific heat data for as-grown and OVD samples. The present study reveals that oxygen reduction provides a knob to simultaneously tune the electrical and magnetic properties in metallic $EuTiO_3$. It furthermore lays the foundation for a range of queries, such as searches for a multiferroic metal and superconductivity, the further fine-tuning of the carrier concentration and magnetic order, and the systematic study of the interplay between the incommensurate structural instability, antiferrodistortive modes, and oxygen vacancies and other defects.

## Acknowledgments

We thank D. Pelc, S. Sharma, Y. Tao, and C. Leighton for discussions, and Y. Tao and C. Leighton for their help with the specific heat measurements. This work was supported by the Department of Energy through the University of Minnesota Center for Quantum Materials, under grant number DE-SC-0016371. Research conducted at the Center for High-Energy x-ray Science (CHEXS) is supported by the National Science Foundation (BIO, ENG, and MPS Directorates) under award DMR-2342336.

## References

[1] M. J. Coak, C. R. S. Haines, C. Liu, S. E. Rowley, G. G. Lonzarich, and S. S. Saxena, Quantum critical phenomena in a compressible displacive ferroelectric, Proceedings of the National Academy of Sciences **117**, 12707 (2020).

[2] S. E. Rowley, L. J. Spalek, R. P. Smith, M. P. M. Dean, M. Itoh, J. F. Scott, G. G. Lonzarich, and S. S. Saxena, Ferroelectric quantum criticality, Nature Phys **10**, 367 (2014).

[3] K. A. Müller and H. Burkard, $SrTiO_3$: An intrinsic quantum paraelectric below 4 K, Phys. Rev. B **19**, 3593 (1979).

[4] X. Li, B. Fauqué, Z. Zhu, and K. Behnia, Phonon thermal hall effect in strontium titanate, Phys. Rev. Lett. **124**, 105901 (2020).

[5] A. Narayan, A. Cano, A. V. Balatsky, and N. A. Spaldin, Multiferroic quantum criticality, Nature Mater **18**, 3 (2019).

[6] Y. Yamada and G. Shirane, Neutron scattering and nature of the soft optical phonon in $SrTiO_3$, J. Phys. Soc. Jpn. **26**, 396 (1969).

[7] T. Esswein and N. A. Spaldin, Ferroelectric, quantum paraelectric, or paraelectric? Calculating the evolution from $BaTiO_3$ to $SrTiO_3$ to $KTaO_3$ using a single-particle quantum mechanical description of the ions, Phys. Rev. Res. **4**, 033020 (2022).

[8] M. N. Gastiasoro, J. Ruhman, and R. M. Fernandes, Superconductivity in dilute $SrTiO_3$: A review, Annals of Physics **417**, 168107 (2020).

[9] X. Lin, G. Bridoux, A. Gourgout, G. Seyfarth, S. Krämer, M. Nardone, B. Fauqué, and K. Behnia, Critical doping for the onset of a two-Band superconducting ground state in $SrTiO_{3-\delta}$, Phys. Rev. Lett. **112**, 207002 (2014).

[10] C. Liu, X. Zhou, D. Hong, B. Fisher, H. Zheng, J. Pearson, J. S. Jiang, D. Jin, M. R. Norman, and A. Bhattacharya, Tunable superconductivity and its origin at $KTaO_3$ interfaces, Nat Commun **14**, 951 (2023).

[11] C. Liu, X. Yan, D. Jin, Y. Ma, H.-W. Hsiao, Y. Lin, T. M. Bretz-Sullivan, X. Zhou, J. Pearson, B. Fisher, J. S. Jiang, W. Han, J.-M. Zuo, J. Wen, D. D. Fong, J. Sun, H. Zhou, and A. Bhattacharya, Two-dimensional superconductivity and anisotropic transport at $KTaO_3$ (111) interfaces, Science **371**, 716 (2021).

[12] M. N. Gastiasoro, M. E. Temperini, P. Barone, and J. Lorenzana, Generalized Rashba electron-phonon coupling and superconductivity in strontium titanate, Phys. Rev. Res. **5**, 023177 (2023).

[13] M. N. Gastiasoro, T. V. Trevisan, and R. M. Fernandes, Anisotropic superconductivity mediated by ferroelectric fluctuations in cubic systems with spin-orbit coupling, Phys. Rev. B **101**, 174501 (2020).

[14] G. Venditti, M. E. Temperini, P. Barone, J. Lorenzana, and M. N. Gastiasoro, Anisotropic Rashba coupling to polar modes in $KTaO_3$, J. Phys. Mater. **6**, 014007 (2023).

[15] K. Dunnett, A. Narayan, N. A. Spaldin, and A. V. Balatsky, Strain and ferroelectric soft-mode induced superconductivity in strontium titanate, Phys. Rev. B **97**, 144506 (2018).

[16] J. M. Edge, Y. Kedem, U. Aschauer, N. A. Spaldin, and A. V. Balatsky, Quantum critical origin of the superconducting dome in $SrTiO_3$, Phys. Rev. Lett. **115**, 247002 (2015).

[17] C. W. Rischau, X. Lin, C. P. Grams, D. Finck, S. Harms, J. Engelmayer, T. Lorenz, Y. Gallais, B. Fauqué, J. Hemberger, and K. Behnia, A ferroelectric quantum phase transition inside the superconducting dome of $Sr_{1-x}Ca_xTiO_{3-\delta}$, Nature Phys **13**, 643 (2017).

[18] Y. Tomioka, N. Shirakawa, and I. H. Inoue, Superconductivity enhancement in polar metal regions of $Sr_{0.95}Ba_{0.05}TiO_3$ and $Sr_{0.985}Ca_{0.015}TiO_3$ revealed by systematic Nb doping, Npj Quantum Mater. **7**, 1 (2022).
[19] A. Stucky, G. W. Scheerer, Z. Ren, D. Jaccard, J.-M. Poumirol, C. Barreteau, E. Giannini, and D. van der Marel, Isotope effect in superconducting n-doped $SrTiO_3$, Sci Rep **6**, 37582 (2016).
[20] Y. Tomioka, N. Shirakawa, K. Shibuya, and I. H. Inoue, Enhanced superconductivity close to a non-magnetic quantum critical point in electron-doped strontium titanate, Nat Commun **10**, 738 (2019).
[21] C. Herrera, J. Cerbin, A. Jayakody, K. Dunnett, A. V. Balatsky, and I. Sochnikov, Strain-engineered interaction of quantum polar and superconducting phases, Phys. Rev. Mater. **3**, 124801 (2019).
[22] K. Ahadi, L. Galletti, Y. Li, S. Salmani-Rezaie, W. Wu, and S. Stemmer, Enhancing superconductivity in $SrTiO_3$ films with strain, Sci. Adv. **5**, eaaw0120 (2019).
[23] D. J. Scalapino, A common thread: The pairing interaction for unconventional superconductors, Rev. Mod. Phys. **84**, 1383 (2012).
[24] P. Gegenwart, Q. Si, and F. Steglich, Quantum criticality in heavy-fermion metals, Nature Phys **4**, 186 (2008).
[25] J. H. Lee, X. Ke, N. J. Podraza, L. F. Kourkoutis, T. Heeg, M. Roeckerath, J. W. Freeland, C. J. Fennie, J. Schubert, D. A. Muller, P. Schiffer, and D. G. Schlom, Optical band gap and magnetic properties of unstrained $EuTiO_3$ films, Applied Physics Letters **94**, 212509 (2009).
[26] A. Jaoui, S. Jiang, X. Li, Y. Tomioka, I. H. Inoue, J. Engelmayer, R. Sharma, L. Pätzold, T. Lorenz, B. Fauqué, and K. Behnia, Glasslike thermal conductivity and narrow insulating gap of $EuTiO_3$, Phys. Rev. Mater. **7**, 094604 (2023).
[27] O. I. Malyi, X.-G. Zhao, A. Bussmann-Holder, and A. Zunger, Local positional and spin symmetry breaking as a source of magnetism and insulation in paramagnetic $EuTiO_3$, Phys. Rev. Mater. **6**, 034604 (2022).
[28] H. Akamatsu, K. Fujita, H. Hayashi, T. Kawamoto, Y. Kumagai, Y. Zong, K. Iwata, F. Oba, I. Tanaka, and K. Tanaka, Crystal and electronic structure and magnetic properties of divalent europium perovskite oxides $EuMO_3$ (M = Ti, Zr, and Hf): experimental and first-principles approaches, Inorg. Chem. **51**, 4560 (2012).
[29] V. Goian, S. Kamba, J. Hlinka, P. Vaněk, A. A. Belik, T. Kolodiazhnyi, and J. Petzelt, Polar phonon mixing in magnetoelectric $EuTiO_3$, Eur. Phys. J. B **71**, 429 (2009).
[30] S. Kamba, D. Nuzhnyy, P. Vaněk, M. Savinov, K. Knížek, Z. Shen, E. Šantavá, K. Maca, M. Sadowski, and J. Petzelt, Magnetodielectric effect and optic soft mode behaviour in quantum paraelectric $EuTiO_3$ ceramics, EPL **80**, 27002 (2007).
[31] D. S. Ellis, H. Uchiyama, S. Tsutsui, K. Sugimoto, K. Kato, D. Ishikawa, and A. Q. R. Baron, Phonon softening and dispersion in $EuTiO_3$, Phys. Rev. B **86**, 220301 (2012).
[32] V. Scagnoli, M. Allieta, H. Walker, M. Scavini, T. Katsufuji, L. Sagarna, O. Zaharko, and C. Mazzoli, $EuTiO_3$ magnetic structure studied by neutron powder diffraction and resonant x-ray scattering, Phys. Rev. B **86**, 094432 (2012).
[33] T. R. McGuire, M. W. Shafer, R. J. Joenk, H. A. Alperin, and S. J. Pickart, Magnetic structure of $EuTiO_3$, Journal of Applied Physics **37**, 981 (1966).

[34] D. Repček, P. Proschek, M. Savinov, M. Kachlík, J. Pospíšil, J. Drahokoupil, P. Doležal, J. Prokleška, and S. Kamba, Multiferroic quantum criticality in a (Eu,Ba,Sr)$TiO_3$ solid solution, Phys. Rev. B **109**, 214109 (2024).
[35] T. Katsufuji and H. Takagi, Coupling between magnetism and dielectric properties in quantum paraelectric $EuTiO_3$, Phys. Rev. B **64**, 054415 (2001).
[36] J. H. Lee et al., A strong ferroelectric ferromagnet created by means of spin–lattice coupling, Nature **466**, 954 (2010).
[37] C. J. Fennie and K. M. Rabe, Magnetic and electric phase control in epitaxial $EuTiO_3$ from first principles, Phys. Rev. Lett. **97**, 267602 (2006).
[38] X. Ke, T. Birol, R. Misra, J.-H. Lee, B. J. Kirby, D. G. Schlom, C. J. Fennie, and J. W. Freeland, Structural control of magnetic anisotropy in a strain-driven multiferroic $EuTiO_3$ thin film, Phys. Rev. B **88**, 094434 (2013).
[39] T. Birol, N. A. Benedek, H. Das, A. L. Wysocki, A. T. Mulder, B. M. Abbett, E. H. Smith, S. Ghosh, and C. J. Fennie, The magnetoelectric effect in transition metal oxides: Insights and the rational design of new materials from first principles, Current Opinion in Solid State and Materials Science **16**, 227 (2012).
[40] T. Birol and C. J. Fennie, Origin of giant spin-lattice coupling and the suppression of ferroelectricity in $EuTiO_3$ from first principles, Phys. Rev. B **88**, 094103 (2013).
[41] Z. Gui and A. Janotti, Carrier-Density-Induced Ferromagnetism in $EuTiO_3$ Bulk and Heterostructures, Phys. Rev. Lett. **123**, 127201 (2019).
[42] Y. Tomioka, T. Ito, and A. Sawa, Magnetotransport Properties of $Eu_{1-x}La_xTiO_3$ ($0 \leq x \leq 0.07$) Single Crystals, J. Phys. Soc. Jpn. **87**, 094716 (2018).
[43] K. Maruhashi, K. S. Takahashi, M. S. Bahramy, S. Shimizu, R. Kurihara, A. Miyake, M. Tokunaga, Y. Tokura, and M. Kawasaki, Anisotropic quantum transport through a single spin channel in the magnetic semiconductor $EuTiO_3$, Advanced Materials **32**, 1908315 (2020).
[44] K. S. Takahashi, M. Onoda, M. Kawasaki, N. Nagaosa, and Y. Tokura, Control of the anomalous Hall effect by doping in $Eu_{1-x}La_xTiO_3$ thin films, Phys. Rev. Lett. **103**, 057204 (2009).
[45] T. Katsufuji and Y. Tokura, Transport and magnetic properties of a ferromagnetic metal: $Eu_{1-x}R_xTiO_3$, Phys. Rev. B **60**, R15021 (1999).
[46] L. Li, J. R. Morris, M. R. Koehler, Z. Dun, H. Zhou, J. Yan, D. Mandrus, and V. Keppens, Structural and magnetic phase transitions in $EuTi_{1-x}Nb_xO_3$, Phys. Rev. B **92**, 024109 (2015).
[47] D. Akahoshi, S. Koshikawa, T. Nagase, E. Wada, K. Nishina, R. Kajihara, H. Kuwahara, and T. Saito, Magnetic phase diagram for the mixed-valence Eu oxide $EuTi_{1-x}Al_xO_3$, Phys. Rev. B **96**, 184419 (2017).
[48] T. Yamamoto, R. Yoshii, G. Bouilly, Y. Kobayashi, K. Fujita, Y. Kususe, Y. Matsushita, K. Tanaka, and H. Kageyama, An antiferro-to-ferromagnetic transition in EuTiO3–xHx induced by hydride substitution, Inorg. Chem. **54**, 1501 (2015).
[49] K. Maruhashi, K. S. Takahashi, M. S. Bahramy, S. Shimizu, R. Kurihara, A. Miyake, M. Tokunaga, Y. Tokura, and M. Kawasaki, Anisotropic Quantum Transport through a Single Spin Channel in the Magnetic Semiconductor $EuTiO_3$, Advanced Materials **32**, 1908315 (2020).
[50] H. Shin, C. Liu, F. Li, R. Sutarto, B. A. Davidson, and K. Zou, Controlling the electrical and magnetic ground states by doping in the complete phase diagram of titanate $Eu_{1-x}La_xTiO_3$ thin films, Phys. Rev. B **101**, 214105 (2020).

[51] K. Kugimiya, K. Fujita, K. Tanaka, and K. Hirao, Preparation and magnetic properties of oxygen deficient $EuTiO_{3-\delta}$ thin films, Journal of Magnetism and Magnetic Materials **310**, 2268 (2007).

[52] D. Shin, I. Kim, S. Song, Y.-S. Seo, J. Hwang, S. Park, M. Choi, and W. S. Choi, Defect engineering of magnetic ground state in $EuTiO_3$ epitaxial thin films, Journal of the American Ceramic Society **104**, 4606 (2021).

[53] J. Engelmayer, X. Lin, C. P. Grams, R. German, T. Fröhlich, J. Hemberger, K. Behnia, and T. Lorenz, Charge transport in oxygen-deficient $EuTiO_3$: The emerging picture of dilute metallicity in quantum-paraelectric perovskite oxides, Phys. Rev. Mater. **3**, 051401 (2019).

[54] J.-W. Kim, P. Thompson, S. Brown, P. S. Normile, J. A. Schlueter, A. Shkabko, A. Weidenkaff, and P. J. Ryan, Emergent superstructural dynamic order due to competing antiferroelectric and antiferrodistortive instabilities in bulk $EuTiO_3$, Phys. Rev. Lett. **110**, 027201 (2013).

[55] V. Goian, S. Kamba, O. Pacherová, J. Drahokoupil, L. Palatinus, M. Dušek, J. Rohlíček, M. Savinov, F. Laufek, W. Schranz, A. Fuith, M. Kachlík, K. Maca, A. Shkabko, L. Sagarna, A. Weidenkaff, and A. A. Belik, Antiferrodistortive phase transition in $EuTiO_3$, Phys. Rev. B **86**, 054112 (2012).

[56] M. Allieta, M. Scavini, L. J. Spalek, V. Scagnoli, H. C. Walker, C. Panagopoulos, S. S. Saxena, T. Katsufuji, and C. Mazzoli, Role of intrinsic disorder in the structural phase transition of magnetoelectric $EuTiO_3$, Phys. Rev. B **85**, 184107 (2012).

[57] See Supplemental Materials for details regarding sample preparation, measurements, data analysis, and DFT calculations.

[58] G. Kresse and J. Furthmüller, Efficiency of ab-initio total energy calculations for metals and semiconductors using a plane-wave basis set, Computational Materials Science **6**, 15 (1996).

[59] G. Kresse and J. Furthmüller, Efficient iterative schemes for ab initio total-energy calculations using a plane-wave basis set, Phys. Rev. B **54**, 11169 (1996).

[60] P. Giannozzi et al., QUANTUM ESPRESSO: a modular and open-source software project for quantum simulations of materials, J. Phys.: Condens. Matter **21**, 395502 (2009).

[61] P. Giannozzi et al., Advanced capabilities for materials modelling with Quantum ESPRESSO, J. Phys.: Condens. Matter **29**, 465901 (2017).

[62] A. Togo and I. Tanaka, First principles phonon calculations in materials science, Scripta Materialia **108**, 1 (2015).

[63] B. J. Kennedy, G. Murphy, E. Reynolds, M. Avdeev, H. E. R. Brand, and T. Kolodiazhnyi, Studies of the antiferrodistortive transition in $EuTiO_3$, J. Phys.: Condens. Matter **26**, 495901 (2014).

[64] D. Bessas, K. Z. Rushchanskii, M. Kachlik, S. Disch, O. Gourdon, J. Bednarcik, K. Maca, I. Sergueev, S. Kamba, M. Ležaić, and R. P. Hermann, Lattice instabilities in bulk $EuTiO_3$, Phys. Rev. B **88**, 144308 (2013).

[65] G. Shirane and Y. Yamada, Lattice-Dynamical Study of the 110°K Phase Transition in $SrTiO_3$, Phys. Rev. **177**, 858 (1969).

[66] K. Hirota, J. P. Hill, S. M. Shapiro, G. Shirane, and Y. Fujii, Neutron- and x-ray-scattering study of the two length scales in the critical fluctuations of $SrTiO_3$, Phys. Rev. B **52**, 13195 (1995).

[67] K. Z. Rushchanskii, N. A. Spaldin, and M. Ležaić, First-principles prediction of oxygen octahedral rotations in perovskite-structure $EuTiO_3$, Phys. Rev. B **85**, 104109 (2012).

[68] M. Kopecký, J. Fábry, and J. Kub, X-ray diffuse scattering in $SrTiO_3$ and model of atomic displacements, J Appl Cryst **45**, 393 (2012).
[69] M. Zacharias, G. Volonakis, F. Giustino, and J. Even, Anharmonic electron-phonon coupling in ultrasoft and locally disordered perovskites, Npj Comput Mater **9**, 1 (2023).
[70] Z. Zhong, A. Tóth, and K. Held, Theory of spin-orbit coupling at $LaAlO_3/SrTiO_3$ interfaces and $SrTiO_3$ surfaces, Phys. Rev. B **87**, 161102 (2013).
[71] D. van der Marel, J. L. M. van Mechelen, and I. I. Mazin, Common Fermi-liquid origin of $T^2$ resistivity and superconductivity in n-type $SrTiO_3$, Phys. Rev. B **84**, 205111 (2011).
[72] A. Midya, P. Mandal, Km. Rubi, R. Chen, J.-S. Wang, R. Mahendiran, G. Lorusso, and M. Evangelisti, Large adiabatic temperature and magnetic entropy changes in $EuTiO_3$, Phys. Rev. B **93**, 094422 (2016).
[73] A. M. Iqbal and G. H. Jaffari, Effect of stoichiometry on electrical response and polydispersivity related to hopping polarization in $EuTiO_3$, Journal of Applied Physics **125**, 114102 (2019).
[74] L. Li, H. Zhou, J. Yan, D. Mandrus, and V. Keppens, Research Update: Magnetic phase diagram of $EuTi_{1-x}B_xO_3$ (B = Zr, Nb), APL Materials **2**, 110701 (2014).
[75] R. I. Dass and J. B. Goodenough, Itinerant to localized electronic transition in $Sr_2FeMo_{1-x}W_xO_6$, Phys. Rev. B **63**, 064417 (2001).
[76] K.-I. Kobayashi, T. Okuda, Y. Tomioka, T. Kimura, and Y. Tokura, Possible percolation and magnetoresistance in ordered double perovskite alloys $Sr_2Fe(W_{1-x}Mo_x)O_6$, Journal of Magnetism and Magnetic Materials **218**, 17 (2000).
[77] Y. Kususe, H. Murakami, K. Fujita, I. Kakeya, M. Suzuki, S. Murai, and K. Tanaka, Magnetic and transport properties of $EuTiO_3$ thin films doped with Nb, Jpn. J. Appl. Phys. **53**, 05FJ07 (2014).
[78] A. M. Iqbal, G. H. Jaffari, M. Saleemi, and A. Ceylan, Relaxation dynamics and polydispersivity associated with defects and ferroelectric correlations in Ba-doped $EuTiO_3$, J. Phys.: Condens. Matter **29**, 465402 (2017).
[79] B. C. Sales, B. C. Chakoumakos, R. Jin, J. R. Thompson, and D. Mandrus, Structural, magnetic, thermal, and transport properties of $X_8Ga_{16}Ge_{30}$ (X =Eu, Sr, Ba) single crystals, Phys. Rev. B **63**, 245113 (2001).
[80] K. Ahadi, Z. Gui, Z. Porter, J. W. Lynn, Z. Xu, S. D. Wilson, A. Janotti, and S. Stemmer, Carrier density control of magnetism and Berry phases in doped $EuTiO_3$, APL Materials **6**, 056105 (2018).
[81] H. Akamatsu, Y. Kumagai, F. Oba, K. Fujita, H. Murakami, K. Tanaka, and I. Tanaka, Antiferromagnetic superexchange via 3d states of titanium in $EuTiO_3$ as seen from hybrid Hartree-Fock density functional calculations, Phys. Rev. B **83**, 214421 (2011).
[82] P. J. Ryan, J.-W. Kim, T. Birol, P. Thompson, J.-H. Lee, X. Ke, P. S. Normile, E. Karapetrova, P. Schiffer, S. D. Brown, C. J. Fennie, and D. G. Schlom, Reversible control of magnetic interactions by electric field in a single-phase material, Nat Commun **4**, 1334 (2013).
[83] A. Najev, N. Somun, M. Spaić, I. Khayr, M. Greven, A. Klein, M. N. Gastiasoro, and D. Pelc, Electronic spin susceptibility in metallic strontium titanate, Npj Quantum Mater. **10**, 4 (2025).
[84] B. Fauqué, C. Collignon, H. Yoon, Ravi, X. Lin, I. I. Mazin, H. Y. Hwang, and K. Behnia, Electronic band sculpted by oxygen vacancies and indispensable for dilute superconductivity, Phys. Rev. Res. **5**, 033080 (2023).

[85] P. P. Edwards and M. J. Sienko, Universality aspects of the metal-nonmetal transition in condensed media, Phys. Rev. B **17**, 2575 (1978).
[86] A. Spinelli, M. A. Torija, C. Liu, C. Jan, and C. Leighton, Electronic transport in doped $SrTiO_3$: Conduction mechanisms and potential applications, Phys. Rev. B **81**, 155110 (2010).
[87] Y. Huang, Y. Ayino, and B. I. Shklovskii, Metal-insulator transition in n-type bulk crystals and films of strongly compensated $SrTiO_3$, Phys. Rev. Mater. **5**, 044606 (2021).
[88] S. Roy, N. Khan, and P. Mandal, Unconventional transport properties of the itinerant ferromagnet $EuTi_{1-x}Nb_x3$(x=0.10--0.20), Phys. Rev. B **98**, 134428 (2018).
[89] E. McCalla, M. N. Gastiasoro, G. Cassuto, R. M. Fernandes, and C. Leighton, Low-temperature specific heat of doped $SrTiO_3$: Doping dependence of the effective mass and Kadowaki-Woods scaling violation, Phys. Rev. Mater. **3**, 022001 (2019).
[90] N. Choudhury, E. J. Walter, A. I. Kolesnikov, and C.-K. Loong, Large phonon band gap in $SrTiO_3$ and the vibrational signatures of ferroelectricity in $ATiO_3$ perovskites: First-principles lattice dynamics and inelastic neutron scattering, Phys. Rev. B **77**, 134111 (2008).

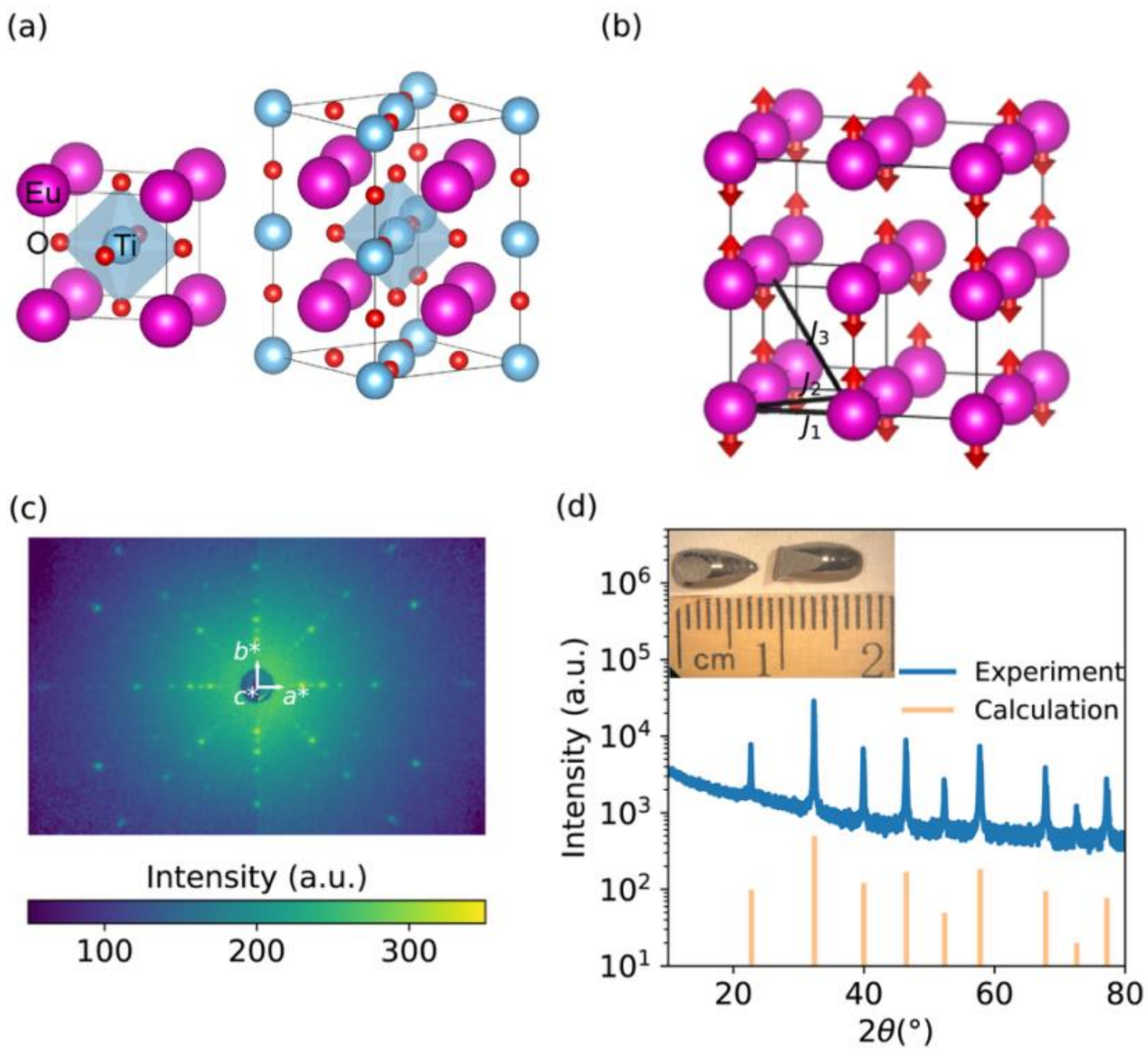

Figure 1. Crystal structure and x-ray characterization of as-grown $EuTiO_3$. (a) Schematic representation of the $Pm\overline{3}m$ cubic and *I4/mcm* tetragonal unit cells. Eu: purple; Ti: blue; O: red. (b) Illustration of *G*-type AFM order in a 2 ×2 ×2 supercell, with a single cubic cell indicated on the bottom left. The nearest, second-nearest, and third-nearest interactions $J_1$, $J_2$, and $J_3$ are indicated by black lines. Ti and O atoms are omitted for clarity. Neighboring spins are antiparallel along <100>. (c) Laue x-ray diffraction pattern obtained with the photon beam along [001], showing the diffraction pattern in the (*HK*0) plane. (d) Room-temperature powder x-ray diffraction result (obtained with Cu K-α photons) compared with calculated Bragg-peak scattering angles and intensities. Inset: two as-grown $EuTiO_3$ crystals, with surfaces polished perpendicular to [100] crystalline direction.

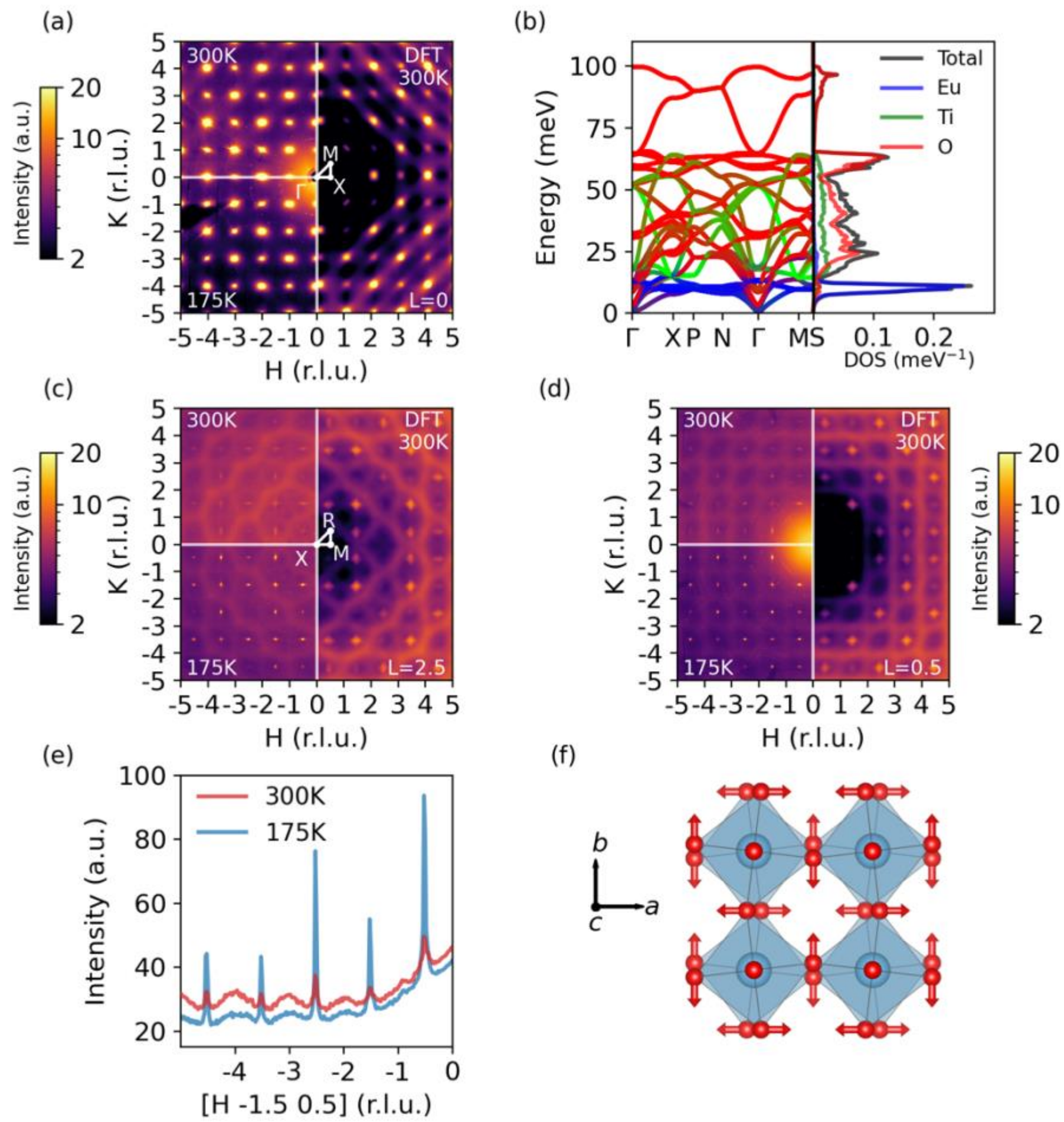

Figure 2. Comparison of synchrotron x-ray diffuse scattering data and first-principles modeling for undoped ETO. (a) X-ray data for the (*HK*0) plane at 300 K and 175 K, above and below the cubic-tetragonal structural transition temperature $T_s \sim 285$ K, compared with calculated TDS. High-symmetry points (in cubic ($Pm\overline{3}m$) notation): Γ (0 0 0), X (0.5 0 0), M (0.5 0.5 0), and (not shown) R (0.5 0.5 0.5). (b) Calculated phonon band structure and density of states, using body-centered tetragonal (*I4/mcm*, $c/a > 1$) notation (we use cubic notation throughout the rest of the paper), with high-symmetry points Γ (0 0 0), X (0 0 0.5), P (0.25 0.25 0.25), N (0 0.5 0), M (0.5 0.5 -0.5), and S (0.374 0.626 -0.374). The phonon dispersion and eigenvectors were used to calculate the dynamical structure factor $S(\mathbf{Q},\omega)$ and the TDS. The different colors indicate the predominant character of the dynamical matrix eigenvectors and not the force constants matrix eigenvectors. Since dynamical matrix eigenvectors include a factor of the square root of masses, the weights of heavier atoms are overestimated. As a result, the acoustic branch near the Γ point, which should have equal weight contributions from all atoms, appears Eu-heavy. (c,d) X-ray data for (*HK*2.5) and (*HK*0.5) planes compared with DFT + *U*. The left panels show the data at 300 K and 175 K, whereas the right panels show the calculated TDS with temperature occupation factor at 300 K. (e) One-dimensional cuts along [*H* -1.5 0.5]. At 300 K, the *R*-points show diffuse scattering associated with the soft-mode due to out-of-phase octahedral rotation. At 175 K, R-point superlattice peaks appear as the system has transitioned into the tetragonal phase. (f) Illustration of phonon eigenvectors and structural distortion associated with the R-point mode.

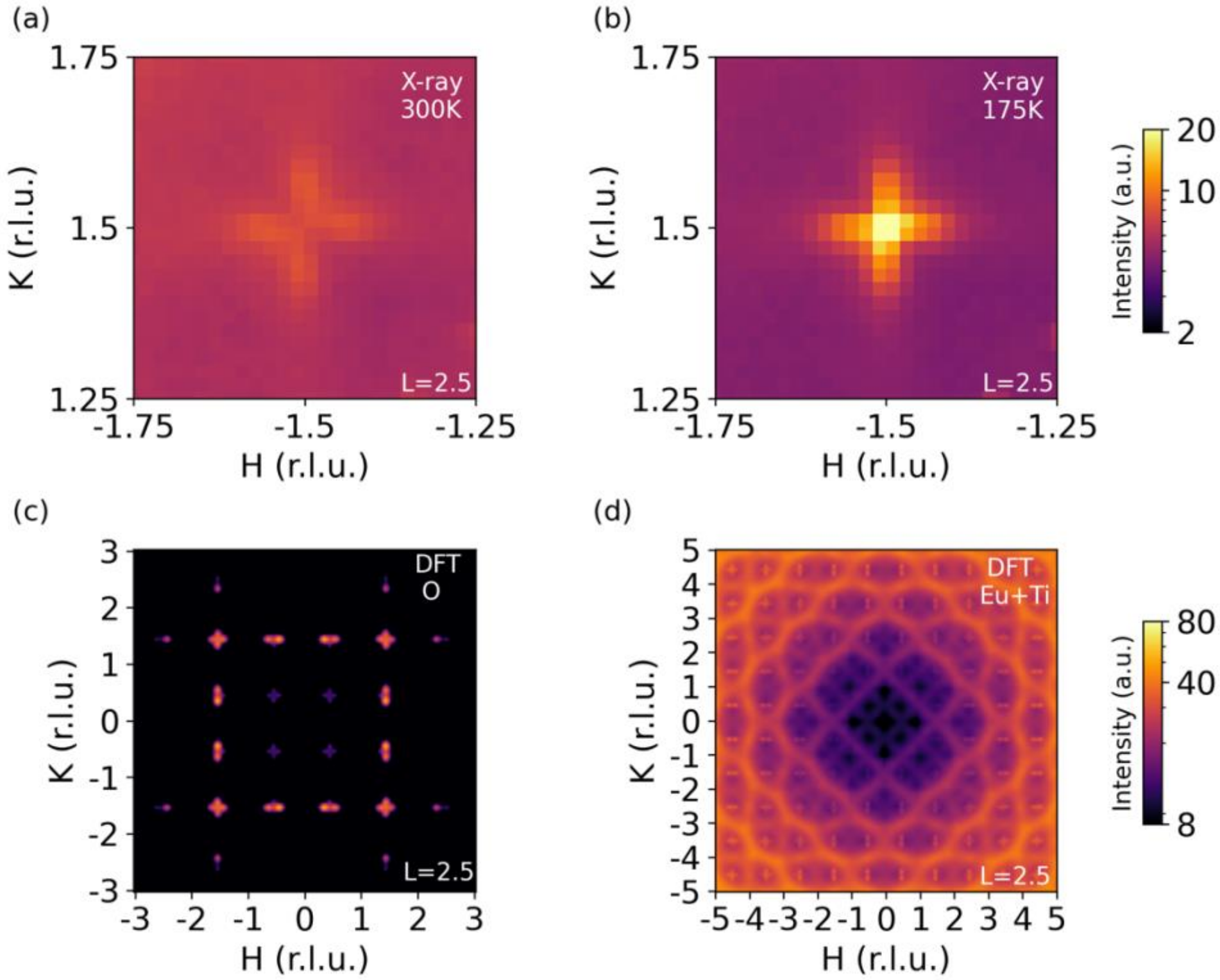

Figure 3. Zoom of synchrotron x-ray diffuse scattering data (same data as in Fig. 2) and element-decomposed thermal diffuse scattering in the (*HK*2.5) plane. The data exhibit cross-shape patterns around R-points at both (a) 300 K and (b) 175 K. (c) The DFT + *U* result of oxygen-only thermal diffuse scattering shows cross-shape patterns at half-integer *H* and *K*. (d) The DFT + *U* result of (Eu + Ti)-decomposed thermal diffuse scattering shows curvilinear features, reflecting valleys and ridges in reciprocal space associated with low-energy phonon modes. The overall agreement between experiment and theory demonstrates that the observed diffuse scattering arises from thermal diffuse scattering rather than static disorder.

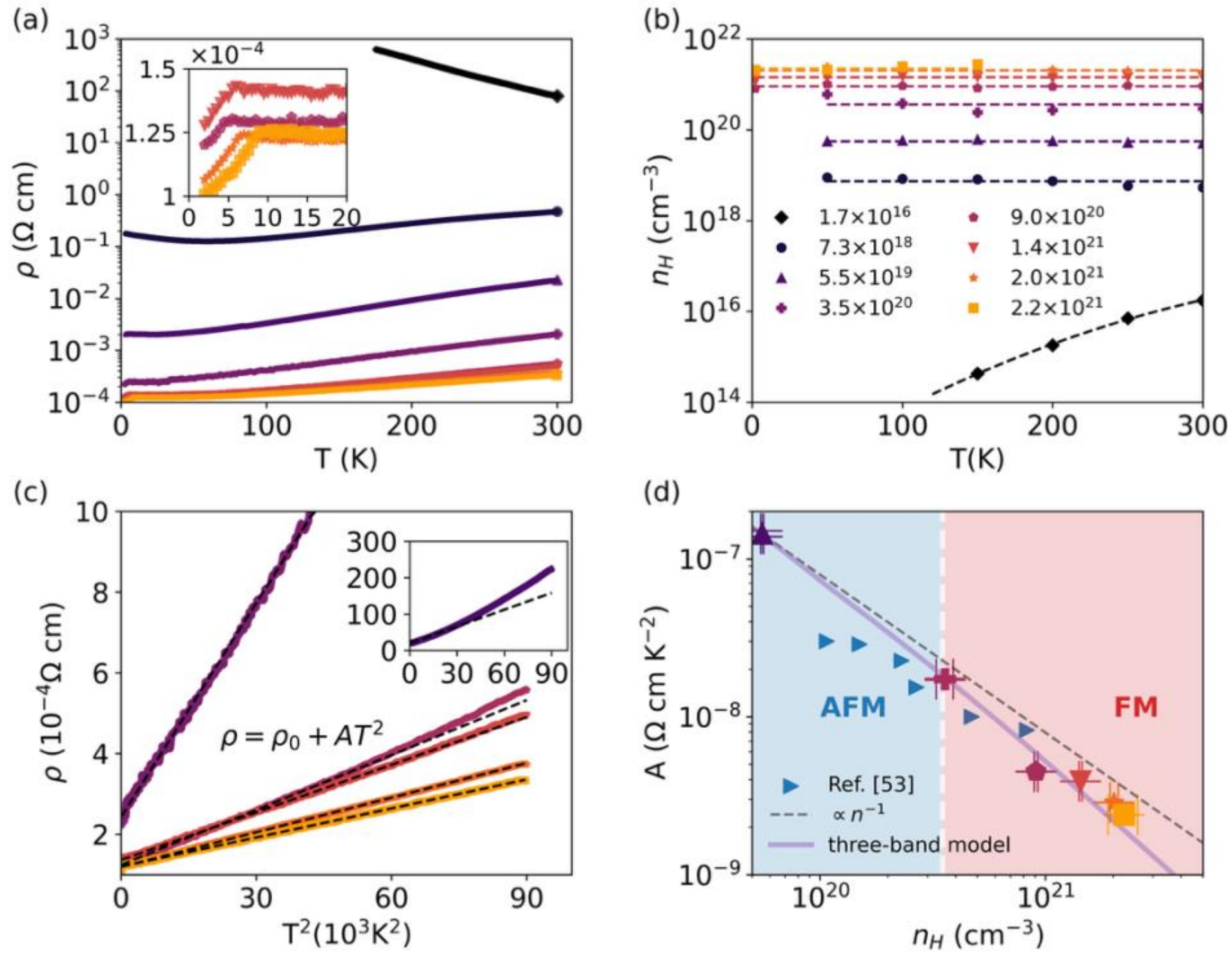

Figure 4. Resistivity and Hall carrier concentration measurements. (a) Resistivity ($\rho$) as a function of temperature ($T$) for varying Hall carrier concentration ($n_H$), with $n_H$ values specified in (b). With increasing doping, the system gradually evolves to a metallic state. Inset: Zoom of low-temperature resistivities of the four samples with the highest carrier concentrations. The cusps indicate a magnetic transition. (b) Carrier concentration as a function of temperature. The as-grown sample exhibits insulating behavior, with a decreasing carrier concentration upon cooling, whereas $n_H$ is nearly temperature-independent for the OVD samples. The specified carrier concentration of the as-grown sample is the room-temperature value, whereas in all other cases it is the temperature-average. (c) $\rho$ vs. $T^2$ and results of fits (dashed lines) to $\rho = \rho_0 + AT^2$ for the six highest doping levels of the present study ($n_H$ values specified in (b)). Inset: Extended result for $n_H = 5.5 \times 10^{19}$ cm$^{-3}$. (d) Log-log plot of $A(n_H)$. Also shown are prior experimental results [53], the simple heuristic $n^{-1}$ dependence, and the result of our three-band model calculation.

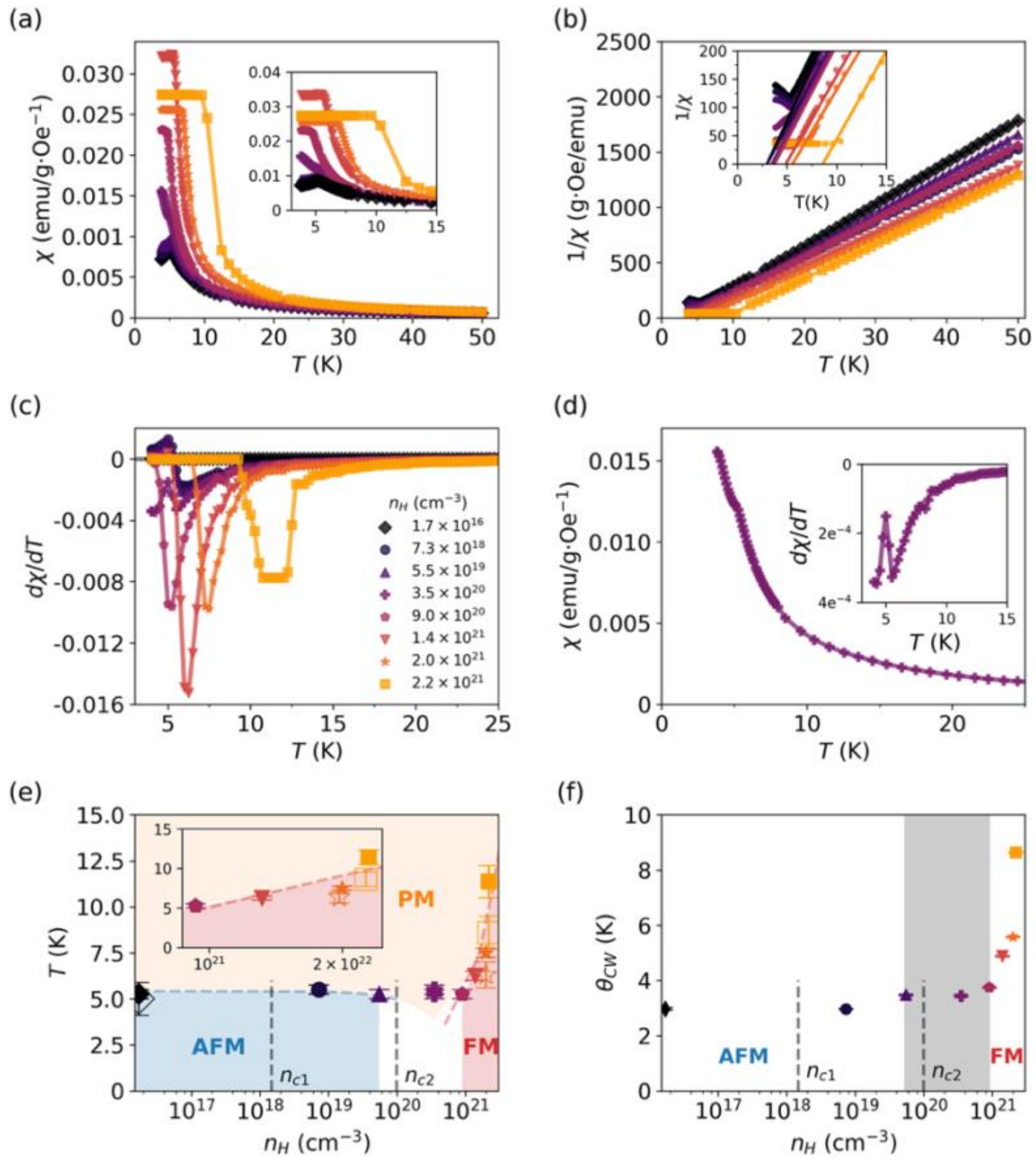

Figure 5. Magnetization measurements and phase diagram of OVD ETO. (a) Temperature dependence of magnetic susceptibility (χ) for samples with varying carrier concentration, obtained with a 20 Oe field under field-cooled conditions along either <100> or <110> crystalline direction. Inset: Zoom of the low-temperature region. The $n_H$ values are indicated in (c). (b) Inverse of magnetic susceptibility as a function of temperature. The inset shows Curie-Weiss fits. (c) Temperature derivative of magnetic susceptibility. (d) χ and d$\chi$/d$T$ for the sample with $n_H \approx 3.5 \times 10^{20}$ cm$^{-3}$; d$\chi$/d$T$ shows a minimum at ~5.5 K. (e) Temperature-doping phase diagram showing the relationship between carrier concentration and magnetic order. The AFM transition temperatures $T_N$ are determined from the cusps in $\chi(T)$ in (a). The FM transition temperatures $T_C$ are determined from the minima of d$\chi(T)$/d$T$ in (c). The sample with $n_H \approx 3.5 \times 10^{20}$ cm$^{-3}$ exhibits the characteristics of a Néel transition, with a kink at $T_N \sim 5.25$ K, as well as a paramagnetic (PM) feature down to the lowest measured temperature of ~ 3.8 K, possibly an indication of macroscopic two-phase co-existence. This sample also shows a local minimum in d$\chi(T)$/d$T$ at $T_C \sim 5.5$ K. The large open symbols indicate the results from specific heat measurements. The white region represents the AFM-FM-PM transition regime, defined by the nearest AFM and FM carrier concentrations around $n_H \approx 3.5 \times 10^{20}$ cm$^{-3}$. Inset: $T_C$ of the four FM samples with a linear carrier-concentration scale. (f) Curie-Weiss temperatures obtained from the fits to the data in (b). In (e) and (f), the Lifshitz transition carrier concentrations $n_{c1}$ and $n_{c2}$, where the Fermi energy crosses the second and third bands, as determined from our three-band model, are indicated as vertical dashed lines. The thick gray band in (f) marks the AFM-FM-PM transition region, as defined same as the white region in (e).

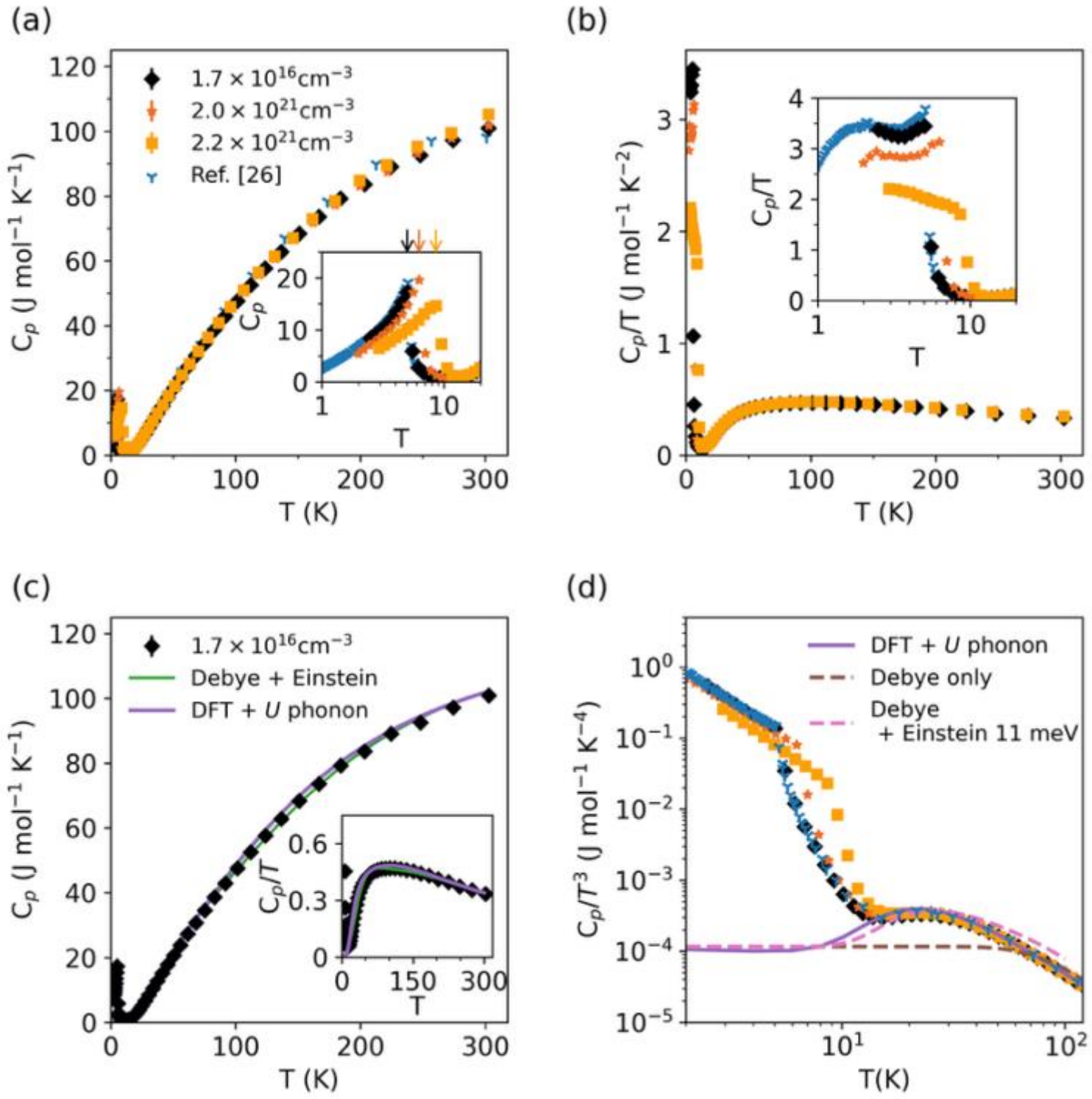

Figure 6. Specific heat measurements. (a) $C_p(T)$ for one as-grown ($n_{\mathrm{H}}$ = 1.7 x $10^{16}$ $\mathrm{cm}^{-3}$) and two FM metallic ($n_H$ = 2.0 x $10^{21}$ $\mathrm{cm}^{-3}$ and 2.2 x $10^{21}$ $\mathrm{cm}^{-3}$) samples compared with a prior result for undoped ETO [26]. The three samples are the same as those for which charge transport and magnetization data are shown in Figs. 3 and 4. Inset: Low-temperature result with logarithmic temperature scale. The downward arrows indicate the peak positions of $C_p(T)$: 5.0 ± 0.9K, 6.3 ± 0.7 K, and 8.6 ± 0.9 K; these characteristic temperatures are slightly lower than $T_{\mathrm{N}}$ and $T_{\mathrm{C}}$ extracted from the magnetization data (see Fig. 5(e)). (b) $C_p/T$ *vs. T*. Inset: Low-temperature results with logarithmic temperature scale. (c) $C_p(T)$ for the undoped sample compared with DFT + $U$ modeling and a Debye-Einstein fit. The modeling includes only the phonon contribution. The Debye-Einstein model gives Debye and Einstein temperatures of $\Theta_{\mathrm{D}} \sim 245$ K and $\Theta_{\mathrm{E}} \sim 635$ K, respectively. Inset: $C_p/T$ *vs. T* compared with DFT+$U$ and Debye-Einstein model. (d) $C_p/T^3$ *vs. T* at low temperatures. All three samples show a hump at about 25 K, consistent with a prior study of undoped ETO *[26]*, which corresponds to the lowest-energy Eu mode in the phonon DOS (see Fig. 2(b)). At low temperatures, $C_p/T^3$ is compared with the DFT + $U$ phonon calculation, a Debye-only fit (resulting in $\Theta_{\mathrm{D}} \sim 436$ K), and a Debye fit plus an Einstein mode at 11 meV.

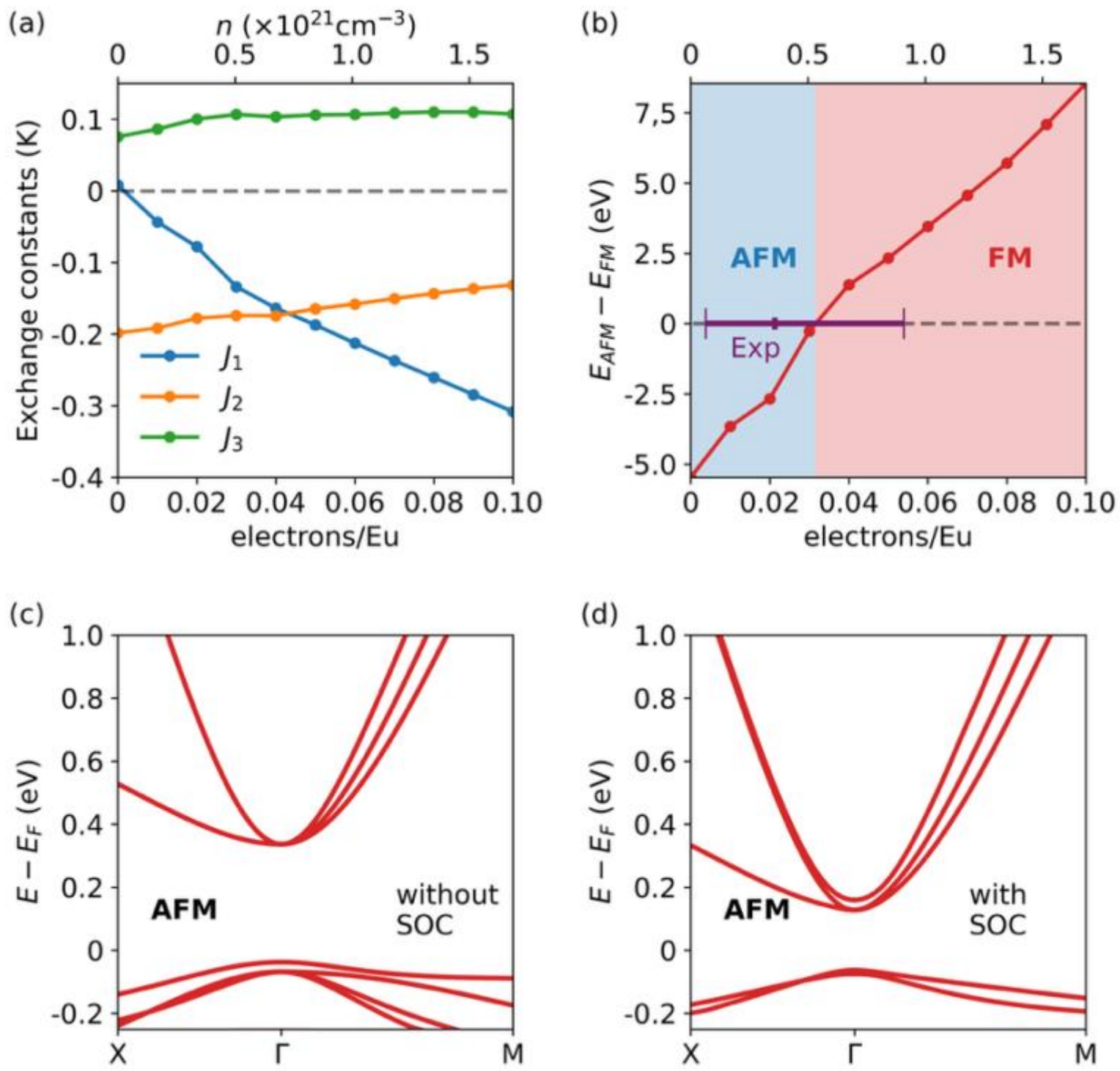

Figure 7. DFT + $U$ results for magnetic exchange constants, difference in ground state energy between the FM and $G$-type AFM states, and band structure. (a) First-, second-, and third-nearest-neighbor exchange interactions vs. electron carrier concentration. Bottom scale: number of electrons per Eu. Top scale: carrier density, denoted here as $n$ (in contrast to the Hall concentration $n_H$ obtained experimentally). (b) Energy difference between AFM and FM states vs. electron carrier concentration. The experimental AFM-FM transition region is indicated as well (see also Fig. 5(e, f)). (c,d) Electronic band structure of undoped cubic ETO in the AFM state with and without spin-orbit coupling (SOC), respectively. SOC lifts the degeneracy at the Γ-point of the three lowest conduction bands. In the AFM state, spin-up and spin-down bands are degenerate. The energy is expressed with respect to the Fermi energy $E_F$.

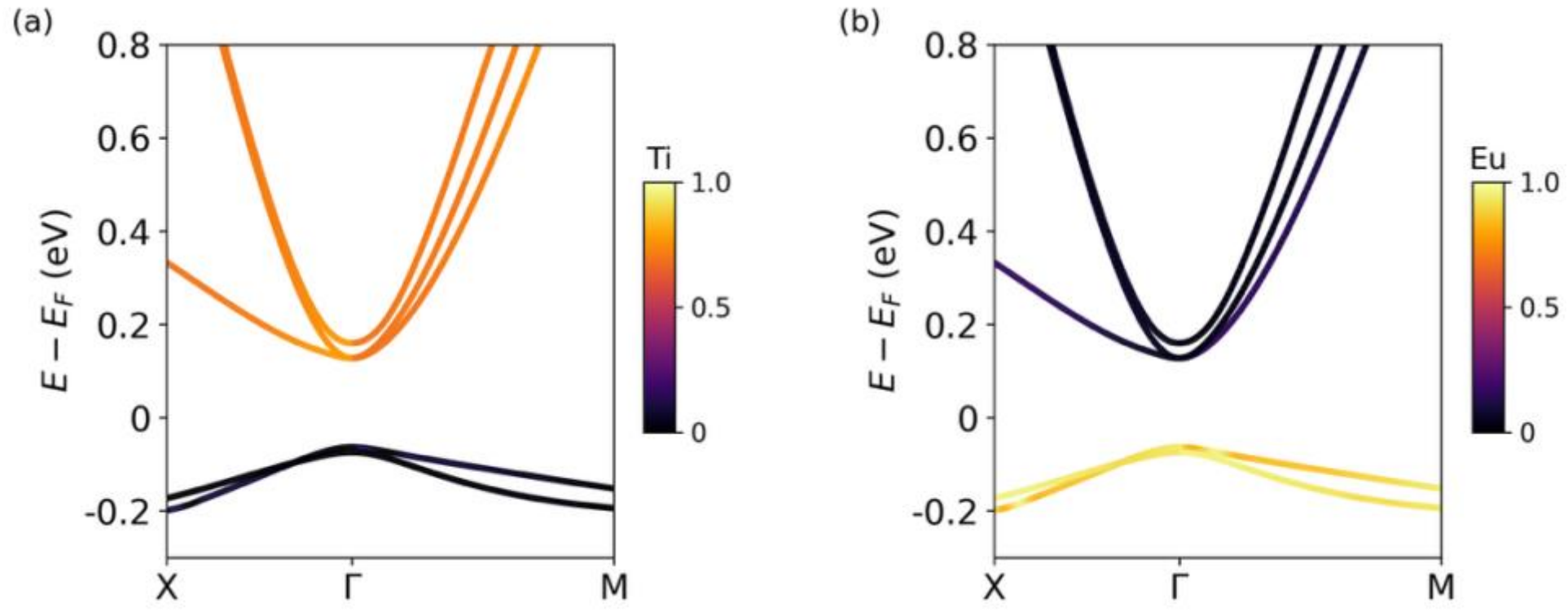

Figure 8. Band structure of undoped AFM ETO with SOC from DFT + $U$ projected to (a) Ti-$d$ and (b) Eu-$f$ orbitals. The energy zero is set to the Fermi energy $E_F$.

**Supplemental Material for**

# Tuning charge-transport properties and magnetic order in metallic $EuTiO_{3-\delta}$

Xing He[1,*,†], Chiou Yang Tan[1], Issam Khayr[1], Zach Van Fossan[2], Richard J. Spieker[1], Dayu Zhai[1], Sarah Anderson[1], Dinesh Shukla[1,3], Suchismita Sarker[4], Javier Garcia-Barriocanal[5], Turan Birol[2], and Martin Greven[1,†]

[1]School of Physics and Astronomy, University of Minnesota, Minneapolis, MN 55455, USA

[2]Department of Chemical Engineering and Materials Science, University of Minnesota, Minneapolis, MN 55455, USA

[3]UGC-DAE Consortium for Scientific Research, Indore, Madhya Pradesh 452001, India

[4]Cornell High Energy Synchrotron Source, Cornell University, Ithaca, NY 14853, USA

[5]Characterization Facility, University of Minnesota, Minneapolis, MN 55455, USA

*Present address: Department of Chemistry, University of Pennsylvania, Philadelphia, PA 19104, USA

[†]Correspondence to: xinghe25@sas.upenn.edu, greven@umn.edu

**Supplementary text: A. Sample preparation. B. In-house x-ray characterization. C. Resistivity and magnetization. D. Synchrotron diffuse x-ray scattering. E. Specific heat. F. Data reproducibility. G. Curie-Weiss fits. H. Spin exchange constants, ground-state energy, and electronic band structure. I. Phonon dispersion, phonon density of states, thermal diffuse scattering. J. Three-band model. K. Additional magnetization and transport data. L. Fits to transport data.**

**Supplementary references S1-21**

**Supplementary figures S1-18**

## A. Sample preparation

$EuTiO_3$ (ETO) crystals were grown with the floating-zone method, as in prior studies [S1,S2]. $Eu_2O_3$ (Thermo Scientific Chemicals, 99.99%) and $TiO_2$ (Sigma-Aldrich, rutile, 99.98%) powders with stoichiometry $Eu_2Ti_2O_7$ were mixed, ground, and then compressed at 60 MPa to form a 6 mm-diameter feed rod of ~8-10 cm length. The feed rod was sintered in a tube furnace at 1050 °C for 10 h, with flowing 5%$H_2$/95%Ar gas to avoid the formation of $Eu_2Ti_2O_7$ with tri-valent $Eu^{3+}$. A small part of each sintered feed rod was used as a seed. Single crystals with a diameter of ~5mm were grown in a 5%$H_2$/95%Ar atmosphere with a growth speed of 10mm/h and a relative rotation of 60 rpm between the feed rod and seed. The atmospheric pressure was maintained at 4-5 bar. A picture of an as-grown crystal is shown in Fig. 1(d) (inset). The crystals were cut into small pieces and characterized with a Laue x-ray diffractometer. Each piece was aligned along <110> or <100> for further measurements. Our reduction technique utilizes $CaH_2$ as a getter material [S3,S4]. As-grown ETO crystals were sealed with $CaH_2$ in a quartz tube under vacuum (pressure below 100 mTorr), with an ETO to $CaH_2$ mass ratio of 1:12. The ETO crystals were situated in a molybdenum boat above the $CaH_2$ to avoid direct contact. The sealed quartz tubes were heated for 12 h at temperatures between 700 °C and 1050 °C, depending on the desired carrier concentration, with higher temperatures resulting in higher carrier concentrations. There exist several possible reduction mechanisms [S5]: $CaH_2$ decomposes and generates $H_2$ gas, and the $H_2$ gas reduces the oxygen from ETO and forms water; $CaH_2$ takes up oxygen and forms $Ca(OH)_2$.

## B. In-house x-ray characterization

Portions of the as-grown crystals were ground for powder x-ray diffraction with a Rigaku Smartlab SE diffractometer, which used Cu K-α photons (λ = 1.5406 Å). Single crystal Laue diffraction patterns such as those in Fig. 1 were taken with a Photonic Science back-reflection system. A typical single-crystal with a polished <100> surface was characterized via single-crystal diffraction with a Rigaku Smartlab XE diffractometer (Cu K-α photons), as shown in Fig. S1.

## C. Resistivity and magnetization

Resistivity and Hall data were obtained with the standard four-point van der Pauw method using a Quantum Design Inc. Physical Property Measurement System (PPMS) in AC drive mode. The four contacts were made with a wire bonder and aluminum wires were used. For metallic samples, the typical contact resistance was 1 Ω. The resistivity and Hall data are shown in Figs. 3 and 4.

Magnetization measurements were performed with a Quantum Design Inc. Magnetic Property Measurement System (MPMS) in a 20 Oe DC field. The field was applied at 50 K, and data were collected upon cooling to about 3.8 K. The magnetization data are shown in Fig. 5.

## D. Synchrotron diffuse x-ray scattering

Synchrotron diffuse x-ray scattering measurements was performed at beamline QM$^2$ ID4B at the Cornell High Energy Synchrotron Source (CHESS). The incident photon energy was 50 keV. Three-axis rotation was used to cover a large $\boldsymbol{Q}$-space volume. An Oxford Cryosystems $N_2$ Cryostream was used to cool the sample. Date for an as-grown sample, measured at 300 K and 175 K, are shown in Figs. 2 and 3. We also measured the highest doped sample ($n_H$ = 2.2 x $10^{21}$ $cm^{-3}$) at 300 K, and the diffuse scattering pattern is shown in Fig. S3. The doped sample exhibits features that are similar to those for undoped ETO, including sharp Bragg peaks in the *HK*0 plane (Fig. S3(a)) indicative of high crystalline quality, and thermal diffuse scattering with streaks and curvilinear features in the half-integer planes *HK*0.5 (Fig. S3(b)) and *HK*2.5 (Fig. S3(c)). However, we note that the doped sample does not show diffuse features at *R* points, likely because of the lower antiferrodistortive (AFD) transition temperature [S6] and the resultant stiffer room-temperature *R*-point soft mode, similar to OVD $SrTiO_3$ [S7]. We did not determine the doping dependence of the AFD transition temperature in the present study.

### E. Specific heat

We used relaxation calorimetry in a PPMS to measure the specific heat ($C_p$) of an as-grown sample that exhibits AFM order ($n_H$ = 1.7 x $10^{16}$ $cm^{-3}$ at 300 K) and of two OVD samples that exhibit FM order ($n_H$ = 2.0 x $10^{21}$ $cm^{-3}$ and 2.2 x $10^{21}$ $cm^{-3}$). The data are shown in Fig. 6. Apiezon N grease was used to affix crystals to PPMS platforms and 2% temperature pulses were used. The measurements were conducted between 300 K and 2 K and performed three times at each temperature for improved accuracy. The data for the undoped sample were fit between 12 and 300 K to a standard Debye-Einstein model [S8]:

$$C_p(T) = 9(1-X)NN_Ak_B\left(\frac{T}{\Theta_D}\right)^3\int_0^{\Theta_D/T}\frac{e^xx^4}{(e^x-1)^2}dx + 3XNN_Ak_B\frac{e^{\Theta_E/T}}{(e^{\Theta_E/T}-1)^2}(\Theta_E/T)^2 \quad \text{(S1)}$$

where $X$ and 1 - $X$ are the respective weights of the Einstein and Debye contributions, $N$ is the number of atoms in the unit cell, $N_A$ is Avogadro's number, $k_B$ is Boltzmann's constant, and $x = \hbar\omega/k_BT$, where $\omega$ is the phonon frequency. The fit results for the Debye and Einstein temperatures are $\Theta_D$ = 245(10) K and $\Theta_E$ = 635(18) K, respectively. We also estimated $C_p$ from our DFT + $U$ phonon calculation by using the equation:

$$C_p = \sum_{qv} k_B\left(\frac{\hbar\omega(qv)}{k_BT}\right)^2\frac{exp(\hbar\omega(qv)/k_BT)}{[exp(\hbar\omega(qv)/k_BT)-1]^2} \quad \text{(S2)}$$

where $q$ is the wavevector, $v$ is the mode, and $\omega$ is the phonon frequency.

### F. Data reproducibility

Samples reduced under the same nominal conditions generally exhibited closely similar resistivities and magnetic susceptibilities, as demonstrated in Figs. S3 and S4.

### G. Curie-Weiss fits

The magnetic susceptibility was fit to the Curie-Weiss form C = $\chi/(T - \theta_{CW})$, and the fit results and extracted Curie temperatures $\theta_{CW}$ are summarized in Figs. 4 and S5. As $EuTiO_{3-\delta}$ transitions from the AFM to the FM state, $\theta_{CW}$ is seen to increase. Similar to our analysis of the resistivity data (see below), we investigated how the lower limit $T_{min}$ of the fit temperature interval [$T_{min}$, 50 K] affects the extracted $\theta_{CW}$ and the $R^2$ value. By varying $T_{min}$ from 10 K to 20 K, we obtained the error estimates in Figs. 4(f) and S5(b). The corresponding $R^2$ values are shown in Fig. S5(c).

We also considered if the sample with the Hall concentration $n_H$ = 3.5 x $10^{20}$ $cm^{-3}$ resides in a miscibility gap, i.e., if its response reflects a mixed state associated with adjacent doping levels. To this end, we modeled this sample's susceptibility as a linear superposition of the susceptibilities of the two measured samples with adjacent doping levels, $n_H$ = 5.5 x $10^{19}$ $cm^{-3}$ and 9.0 x $10^{20}$ $cm^{-3}$, $\chi$ (3.5 x $10^{20}$ $cm^{-3}$) = $Coef_1\chi$(5.5 x $10^{19}$ $cm^{-3}$) + $Coef_2\chi$(9.0 x $10^{20}$ $cm^{-3}$), as shown in Fig. S6 (blue dashed curve). The fit values of coefficients are $Coef_1 \approx 0.65$ and $Coef_2 \approx 0.35$. This superposition captures the response above the magnetic transition temperature, but it does not describe the paramagnetic trend at low temperatures. Adding an extra paramagnetic contribution, $\chi = C/T$ (orange curve in Fig. S6), approximately captures the low-temperature trend (green dashed curve in Fig. S6). As described in the main text, this can be understood within a three-dimensional percolation picture.

### H. Spin exchange constants, ground state energy, and electronic band structure

Density functional theory (DFT) simulations were performed with VASP [S9,S10] using the Projector-Augmented Wave (PAW) method and Perdew–Burke–Ernzerhof (PBE) Generalized Gradient Approximation (GGA) [S11,S12] exchange correlation functional. In order to account for the strong interactions between Eu *f*-electrons, the DFT + $U$ formalism was used with the on-site Coulomb interaction set to $U_{Eu}$ = 5.7 eV for calculations related to the magnetic exchange constants, while values of 5.7 eV and 6.2 eV were used for electronic band structure calculations. The experimentally observed room-temperature lattice constant of about 3.90 Å [S13] was used for all calculations, along with a plane-wave cutoff energy of 600 eV, on-site exchange of $J_{Eu}$ = 1.0 eV, and a 6 x 6 x 6 *k*-point mesh.

For the determination of the magnetic exchange constants, electronic relaxation was performed for multiple collinear magnetic orders in a 2x2x2 40-atom supercell. The Heisenberg model was then used to perform a linear regression in which the coefficients associated with first-, second-, and third-nearest-neighbor exchange interactions $J_1$, $J_2$, and $J_3$ were determined. The electronic band structure was obtained using a two-step procedure in which a self-consistent electronic relaxation was performed for a cubic 2x2x2 supercell with spin-orbit coupling, followed by a non-self-consistent electronic relaxation along desired high symmetry directions in the Brillouin zone using the converged charge density. The bands of the primitive 5-atom unit cell were identified from the band structure of the 2x2x2 supercell using the unfolding method provided by the PyProcar software package [S14]. While a value of $U_{Eu}$ = 5.7 eV offers experimental agreement with the ratio of the Curie and Neel temperatures of cubic ETO, this was found to cause unwanted hybridization between the Eu-4$f$ and Ti-3$d$ bands, which makes the system metallic. A value of $U_{\mathrm{Eu}}$ = 6.2 eV was therefore used to remove the hybridization between the Eu-4$f$ and Ti-3$d$ bands and create a gap which would allow the use of the three-band model considered in this study. Given the close similarity between the curvature of the Ti-3$d$ bands for the two different values of $U_{\mathrm{Eu}}$, we expect the three-band model fitted using $U_{\mathrm{Eu}}$ = 6.2 eV to retain interpretability. The calculated unfolded AFM band structures using $U_{\mathrm{Eu}}$ = 5.7 eV and 6.2 eV without and with SOC are shown in Fig. S7. The calculated FM band structures using $U_{\mathrm{Eu}}$ = 5.7 eV and 6.2 eV without and with SOC are shown in Figs. S8 and S9.

## I. Phonon dispersion, phonon density of states, thermal diffuse scattering

DFT simulations were performed using Quantum Espresso (QE) [S15,S16] for phonon property simulations. A 6 x 6 x 5 Monkhorst-Pack electronic $k$-point mesh and plane-wave cutoff energy of 480 Ry were used for a tetragonal unit cell (20 atoms). The generalized gradient approximation (GGA) in the Perdew–Burke–Ernzerhof revised for solid (PBE-sol) was utilized. We used the DFT + $U$ formalism with an on-site Coulomb parameter $U$ = 8 eV and Hund's exchange $J_H$ = 1.0 eV. The lattice constant of the tetragonal cell was fully relaxed with energy criteria of $10^{-5}$ Ry and force criteria of $10^{-4}$ Ry/Bohr (relaxed lattice constants: $a = b$ = 5.487 Å, c = 7.823 Å are close to the experimental lattice constants at 175 K: $a = b$ = 5.515 Å, c = 7.805 Å) [S13]. Harmonic phonon calculations were performed using the nonzero-displacement approach as implemented in the Phonopy package [S17]. A 2 x 2 x 1 supercell, a 3 x 3 x 5 Monkhorst-Pack electronic $k$-point mesh, and a plane-wave cutoff energy of 480 Ry were used in QE. The phonon calculation was also used to compare with specific heat measurements. We also calculated the phonon band structure with $U$ = 6.8 eV, and the obtained dispersions are very similar to those obtained with $U$ = 8.0 eV, as shown in Fig. S10. This indicates that the phonon properties are not very sensitive to the choice of $U$ value. Phonon energy, eigenvectors, and x-ray form factors were used to calculate the dynamic structure factor $S(\mathbf{Q},\omega)$ [S18], which is defined as:

$$\mathrm{S}(\mathbf{Q},\omega) \propto |\mathrm{F}(\mathbf{Q})|^2 \frac{(\mathrm{n}+1)}{\omega} \tag{S3}$$

$$\mathrm{F}(\mathbf{Q}) = \sum_{\mathrm{j}} (\mathbf{Q}\cdot\mathbf{e}_{\mathrm{j}}) \frac{\mathrm{b_j}}{\sqrt{\mathrm{M_j}}} \mathrm{e}^{\mathrm{i}\mathbf{Q}\cdot\mathbf{R}_{\mathrm{j}}}\, \mathrm{e}^{-\mathrm{W_j}} \tag{S4}$$

where $\mathbf{Q}$ is the momentum transfer, $\hbar\omega$ the phonon energy, $\hbar$ the reduced Planck constant, $n$ the Bose-Einstein occupation factor $n = \left[exp\left(\frac{\hbar\omega}{k_B T}\right) - 1\right]^{-1}$, $e_j$ the phonon eigenvector, $b_j$ the x-ray form factor, $j$ the atomic index, $R_j$ the atomic position in the unit cell, and $W_j$ the Debye-Waller factor. The dynamic structure factor was calculated with the Phonopy package [S17], and the thermal diffuse scattering intensity was obtained by integrating $\hbar\omega$ from zero to 30 meV. Eu, Ti, Eu+Ti, and O only diffuse scattering patterns were obtained by setting the x-ray form factors of other elements to be zero, as shown in Figs. S11 and S12.

### J. Three-band model

As discussed in the main text, we adopted the simple parabolic three-band model from Ref. [S6], which studied ETO, but used parameters for $SrTiO_3$, and modified this to the undoped cubic AFM ETO case. Considering the simple case of three parabolic bands with band minima at $E_0$, $E_1$, and $E_2$, and effective masses $m_0$, $m_1$, and $m_2$, the carrier concentration of each band is

$$n_i(E_F) = 1/3\pi \left(\frac{2m_i}{\hbar^2}\right)^{3/2} \int_{E_i}^{E_F} \sqrt{E - E_i}\, dE, \tag{S5}$$

and the total electron density is $n(E_F) = n_0 + n_1 + n_2$. We used $m_0 = 1.24\ m_e$, $m_1 = 3.47\ m_e$, $m_2 = 1.62\ m_e$, $E_0 = 0$ meV, $E_1 = 5$ meV, and $E_2 = 31$ meV from our DFT + $U$ calculation, where $m_e$ is the electron mass. The effective masses were obtained by fitting the lowest three conduction bands along the $\Gamma - X$ direction to a parabolic function,

$$E(k) = E_c + \frac{\hbar^2 k^2}{2m_i}, \tag{S6}$$

where $E(k)$ is the band energy, $k$ the wavevector in units of $Å^{-1}$, $E_c$ the minimum energy, and $m_i$ the effective mass (see Fig. S13). The effective mass $m_i$ is expressed in units of electron mass $m_e$, with $\hbar^2/(2m_e) = 3.81$ eV·Å. We note that for undoped cubic AFM ETO, the band structures, including SOC, show no visible splitting in the two lowest conduction bands, similar to STO [S20]. Including tetragonality in STO results in a small splitting about 5 meV [S21], and thus we approximate the splitting of ETO by $E_1 - E_0 = 5$ meV.

From the inverse function $E_F(n)$, we calculated $A(n) \propto E_F^{-2}(n)$. The exponent $\alpha$ in $A \propto n^{\alpha}$ was obtained from the gradient of a log-log plot of $A(n)$. Figure S14 shows both $A(n)$ and $\alpha(n)$. Similar to the three-band model in Ref. [S6], when the carrier concentration is low, and only one band is filled, $\alpha$ is proportional to $n^{-4/3}$. As the second band starts to be filled, $\alpha$ suddenly increases to $n^{-2/3}$ and gradually reaches $n^{-1}$. In the three-band regime, $n^{-1}$ slowly changes back to $n^{-4/3}$. The Lifshitz transitions occur at the carrier concentrations $n_{c1} = 1.48 \times 10^{18}$ cm$^{-3}$ and $n_{c2} = 9.91 \times 10^{19}$ cm$^{-3}$ (see Fig. S14).

### K. Additional magnetization and transport data

We performed additional $M$ vs. $H$ measurements at 4.2 K of the as-grown and most highly doped samples, as shown in Fig. S15. For both samples, the magnetization gradually saturates at high fields and approaches the ideal $Eu^{2+}$ ($S = 7/2$) spin moment of 7 μB at 5 T.

We performed longitudinal resistance measurements of the $n_H = 2.0 \times 10^{21}$ cm$^{-3}$ sample under different applied magnetic fields, as shown in Fig. S16. As the magnetic field increases, the cusp in the resistance gradually disappears, and the longitudinal resistance decreases, reflecting the large magnetoresistance and the interaction between Eu spins and itinerant Ti 3$d$ electrons.

## L. Fits to transport data

In an effort to evaluate the "optimal" temperature range for fits of the resistivity to the Fermi-liquid relationship $\rho = \rho_0 + AT^2$, fits were performed within varying temperature ranges from 2 K to $T_{max}$ (100K < $T_{max}$ < 300K) for the data obtained at the six highest doping levels. The fit parameter $A$ changes as the fitting range changes, as shown in Fig. S17(b). With increasing doping level, $R^2$ values closest to one were obtained for $T_{max,opt}$ = 150 K, 220 K, 240 K, 280 K, 300 K, and 300 K, as summarized in Figs. S17(c,d). Note that the maximum temperature of our measurements was 300 K, and thus, this analysis is less reliable at the highest doping levels. These $T_{max,opt}$ values were used for the fits in Fig. S17(a) and to obtain $A(n)$. The resultant residual resistivity estimates are shown in Fig. S18.

**Supplemental references**

[S1] D. S. Ellis, H. Uchiyama, S. Tsutsui, K. Sugimoto, K. Kato, D. Ishikawa, and A. Q. R. Baron, Phonon softening and dispersion in $EuTiO_3$, Phys. Rev. B **86**, 220301 (2012).

[S2] A. Midya, P. Mandal, Km. Rubi, R. Chen, J.-S. Wang, R. Mahendiran, G. Lorusso, and M. Evangelisti, Large adiabatic temperature and magnetic entropy changes in $EuTiO_3$, Phys. Rev. B **93**, 094422 (2016).

[S3] D. Li, K. Lee, B. Y. Wang, M. Osada, S. Crossley, H. R. Lee, Y. Cui, Y. Hikita, and H. Y. Hwang, Superconductivity in an infinite-layer nickelate, Nature **572**, 624 (2019).

[S4] J. Zhao, Y.-L. Gan, G. Yang, Y.-G. Zhong, C.-Y. Tang, F.-Z. Yang, G. N. Phan, Q.-T. Sui, Z. Liu, G. Li, X.-G. Qiu, Q.-H. Zhang, J. Shen, T. Qian, L. Lu, L. Yan, G.-D. Gu, and H. Ding, Continuously Doping $BiM_2Sr_2CaCu_2O_{8+\delta}$ into Electron-Doped Superconductor by $CaH_2$ Annealing Method, Chinese Phys. Lett. **39**, 077403 (2022).

[S5] M. A. Hayward, M. A. Green, M. J. Rosseinsky, and J. Sloan, Sodium hydride as a powerful reducing agent for topotactic oxide deintercalation: synthesis and characterization of the nickel(I) oxide $LaNiO_2$, J. Am. Chem. Soc. **121**, 8843 (1999).

[S6] J. Engelmayer, X. Lin, C. P. Grams, R. German, T. Fröhlich, J. Hemberger, K. Behnia, and T. Lorenz, Charge transport in oxygen-deficient $EuTiO_3$: The emerging picture of dilute metallicity in quantum-paraelectric perovskite oxides, Phys. Rev. Mater. **3**, 051401 (2019).

[S7] Q. Tao, B. Loret, B. Xu, X. Yang, C. W. Rischau, X. Lin, B. Fauqué, M. J. Verstraete, and K. Behnia, Nonmonotonic anisotropy in charge conduction induced by antiferrodistortive transition in metallic $SrTiO_3$, Phys. Rev. B **94**, 035111 (2016).

[S8] Y. Zhang, A. Saha, F. Tutt, V. Chaturvedi, B. Voigt, W. Moore, J. Garcia-Barriocanal, T. Birol, and C. Leighton, Thermal properties of the metallic delafossite $PdCoO_2$: A combined experimental and first-principles study, Phys. Rev. Mater. **6**, 115004 (2022).

[S9] G. Kresse and J. Furthmüller, Efficiency of ab-initio total energy calculations for metals and semiconductors using a plane-wave basis set, Computational Materials Science **6**, 15 (1996).

[S10] G. Kresse and J. Furthmüller, Efficient iterative schemes for ab initio total-energy calculations using a plane-wave basis set, Phys. Rev. B **54**, 11169 (1996).

[S11] J. P. Perdew, A. Ruzsinszky, G. I. Csonka, O. A. Vydrov, G. E. Scuseria, L. A. Constantin, X. Zhou, and K. Burke, Restoring the density-gradient expansion for exchange in solids and surfaces, Phys. Rev. Lett. **100**, 136406 (2008).

[S12] P. E. Blöchl, Projector augmented-wave method, Phys. Rev. B **50**, 17953 (1994).

[S13] M. Allieta, M. Scavini, L. J. Spalek, V. Scagnoli, H. C. Walker, C. Panagopoulos, S. S. Saxena, T. Katsufuji, and C. Mazzoli, Role of intrinsic disorder in the structural phase transition of magnetoelectric $EuTiO_3$, Phys. Rev. B **85**, 184107 (2012).

[S14] U. Herath, P. Tavadze, X. He, E. Bousquet, S. Singh, F. Muñoz, and A. H. Romero, PyProcar: A Python library for electronic structure pre/post-processing, Computer Physics Communications **251**, 107080 (2020).

[S15] P. Giannozzi et al., QUANTUM ESPRESSO: a modular and open-source software project for quantum simulations of materials, J. Phys.: Condens. Matter **21**, 395502 (2009).

[S16] P. Giannozzi et al., Advanced capabilities for materials modelling with Quantum ESPRESSO, J. Phys.: Condens. Matter **29**, 465901 (2017).

[S17] A. Togo and I. Tanaka, First principles phonon calculations in materials science, Scripta Materialia **108**, 1 (2015).
[S18] G. L. Squires, *Introduction to the Theory of Thermal Neutron Scattering* (Courier Corporation, 1996).
[S19] K. Maruhashi, K. S. Takahashi, M. S. Bahramy, S. Shimizu, R. Kurihara, A. Miyake, M. Tokunaga, Y. Tokura, and M. Kawasaki, Anisotropic Quantum Transport through a Single Spin Channel in the Magnetic Semiconductor $EuTiO_3$, Advanced Materials **32**, 1908315 (2020).
[S20] Z. Zhong, A. Tóth, and K. Held, Theory of spin-orbit coupling at $LaAlO_3/SrTiO_3$ interfaces and $SrTiO_3$ surfaces, Phys. Rev. B **87**, 161102 (2013).
[S21] D. van der Marel, J. L. M. van Mechelen, and I. I. Mazin, Common Fermi-liquid origin of $T^2$ resistivity and superconductivity in n-type $SrTiO_3$, Phys. Rev. B **84**, 205111 (2011).

## Supplemental figures

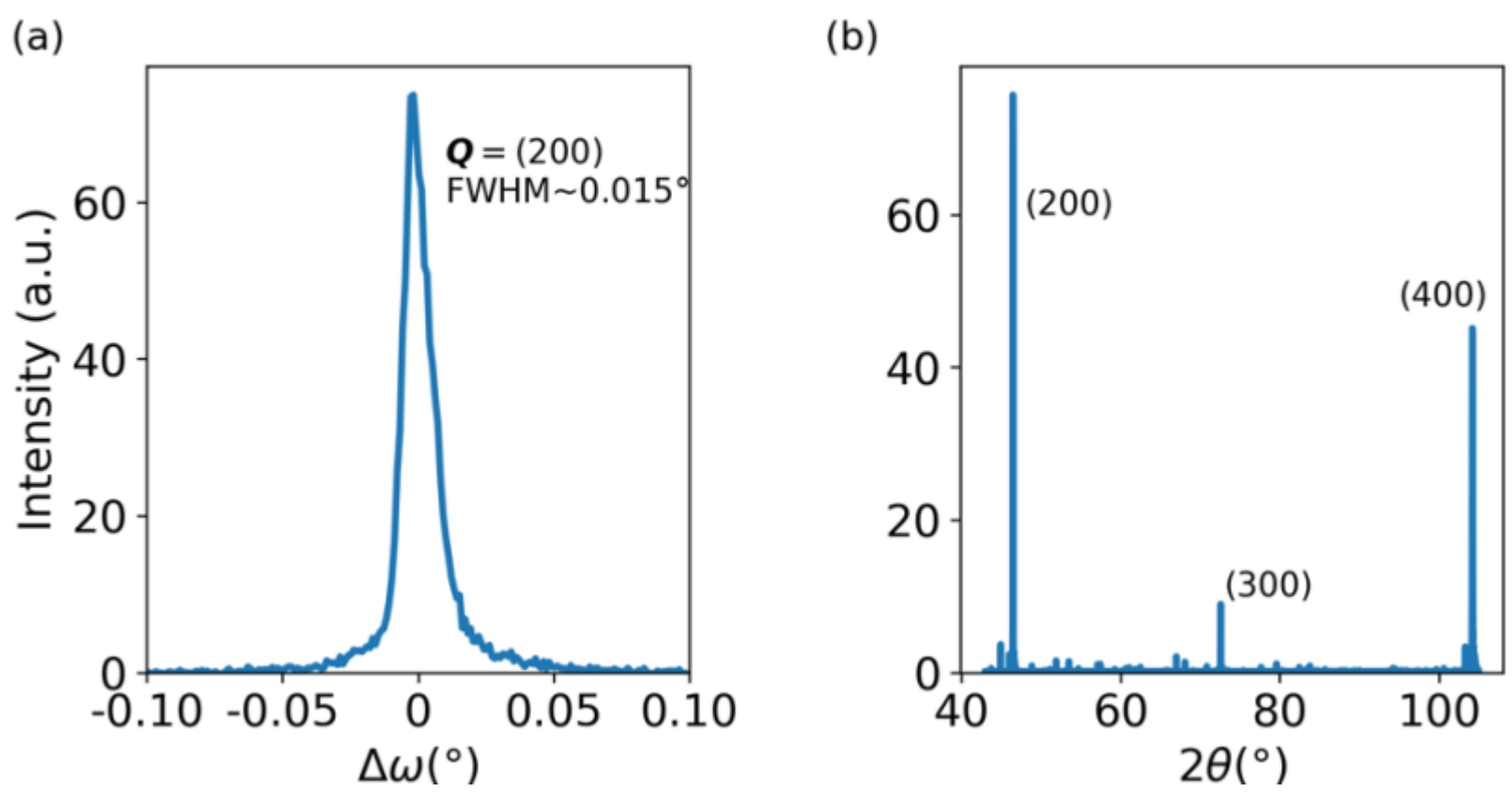

Figure S1. Single-crystal x-ray diffraction result for as-grown $EuTiO_3$. (a) Rocking curve ($\omega$-scan) across the (200) Bragg peak. (b) $\theta/2\theta$ scan along [100].

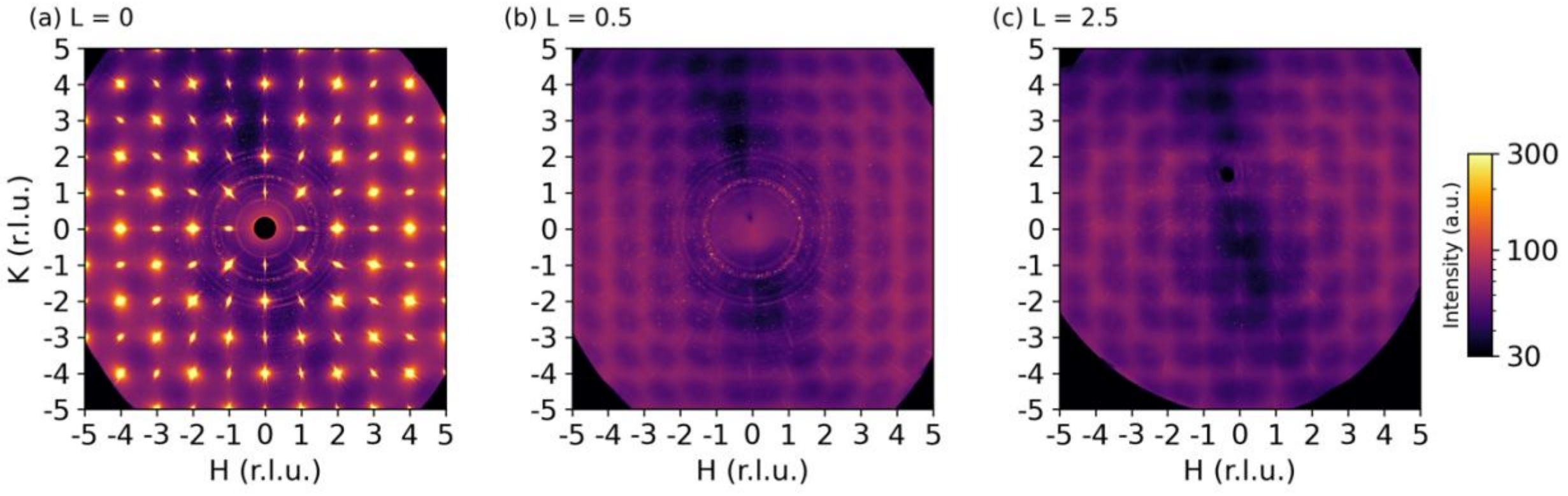

Figure S2. Synchrotron x-ray scattering data for the OVD ETO sample with $n_H = 2.2 \times 10^{21}$ cm$^{-3}$ in the (a) *HK*0, (b) *HK*0.5, and (c) *HK*2.5 planes.

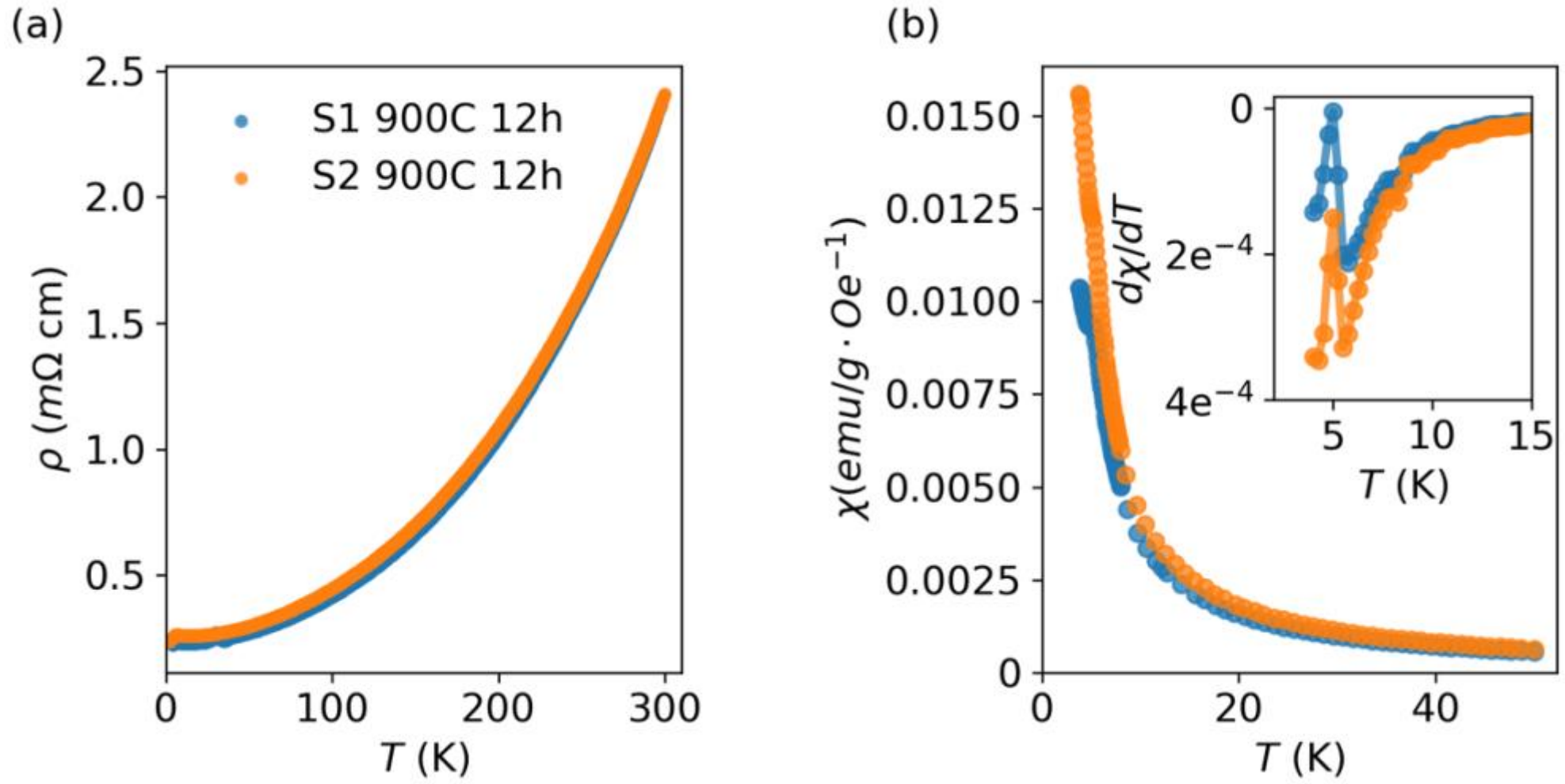

Figure S3. Example of data reproducibility for two samples reduced at 900 °C for 12 h. (a) Resistivity. (b) Corresponding magnetic susceptibility data for the same two samples. The results for sample S2 are the same as those shown in the main text ($n_H = 3.5 \times 10^{20}$ cm$^{-3}$).

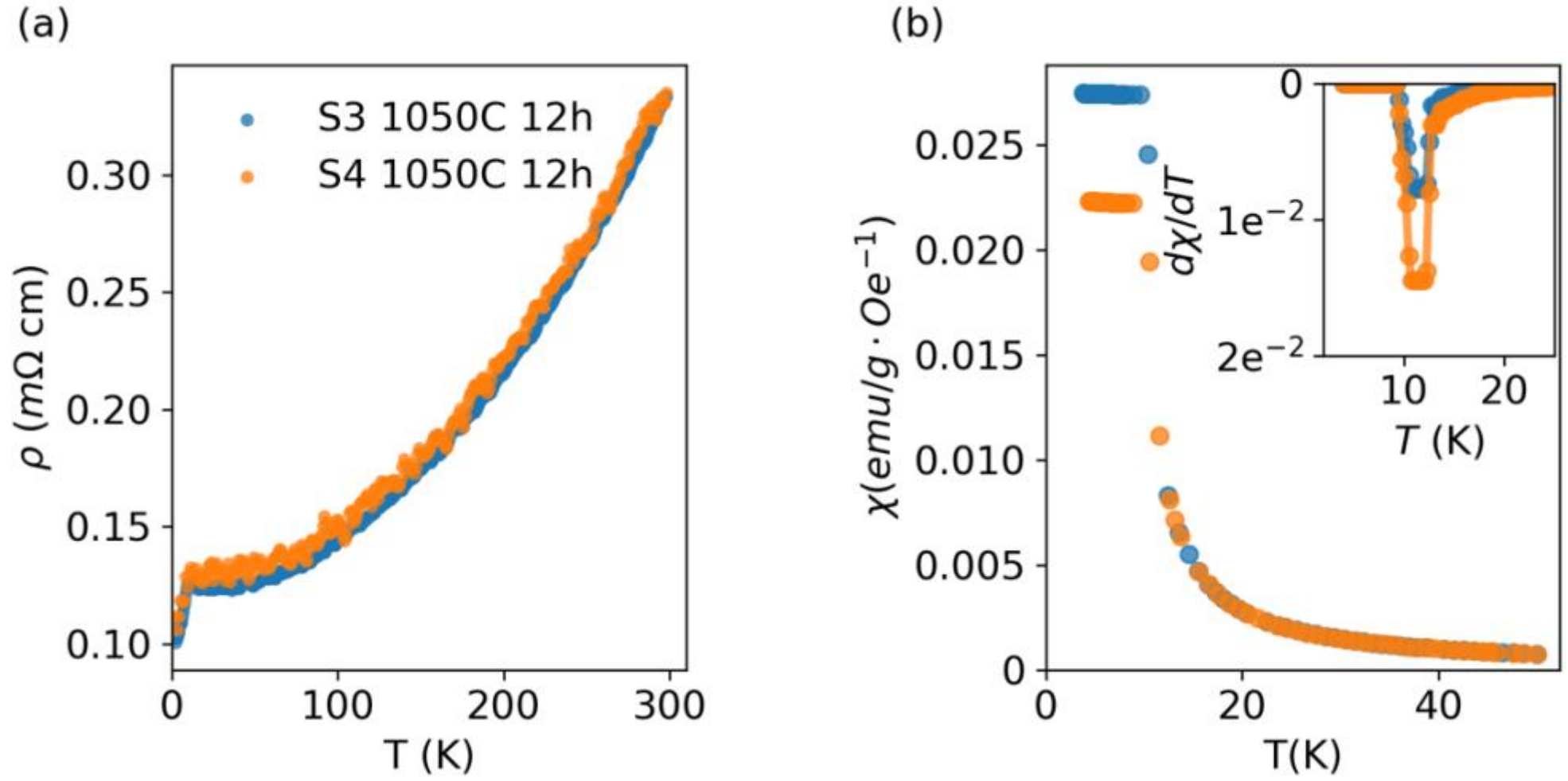

Figure S4. Example of data reproducibility for two samples reduced at 1050 °C for 12 h. (a) Resistivity. (b) Corresponding magnetic susceptibility data. The results for sample S3 are the same as those shown in the main text ($n_H = 2.2 \times 10^{21}$ cm$^{-3}$).

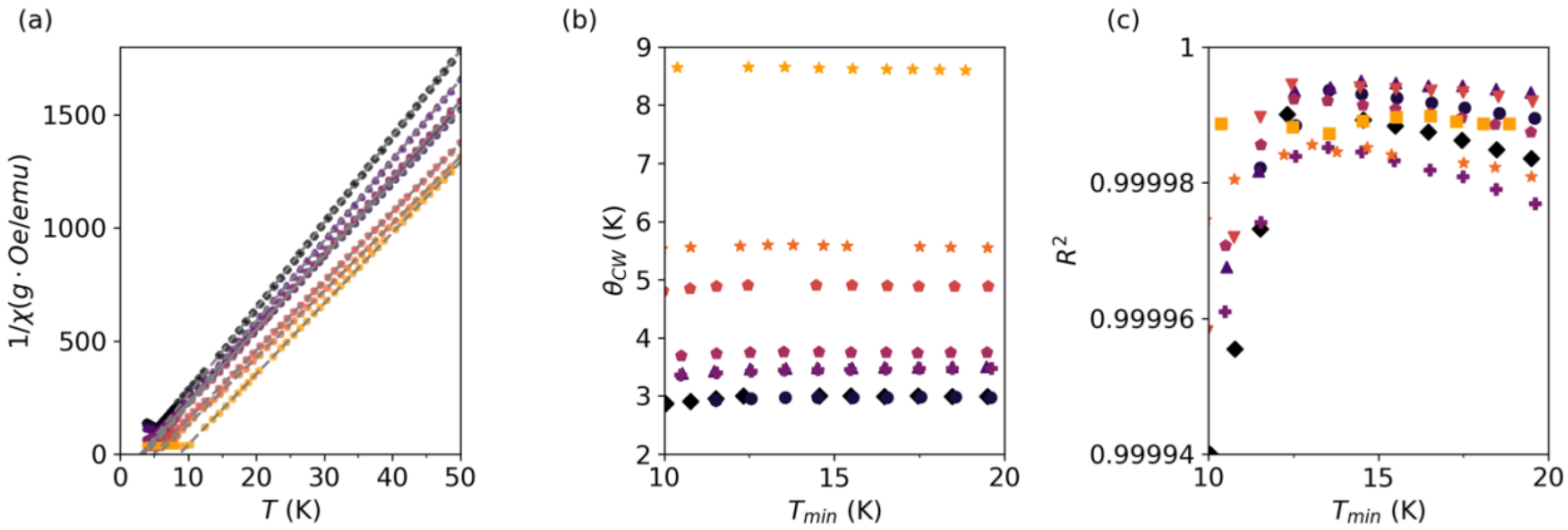

Figure S5. Curie-Weiss fit results. (a) Data for $1/\chi$ and fits (dashed lines, fit range 12.5 - 50 K). The colors and the markers are the same as the ones in Figs. 3, 4, and 5. (b) Fit results for $\theta_{CW}$ for different fit interval cut-offs $T_{\min}$ in the 10 K to 20 K range. (c) $R^2$ values corresponding to (b).

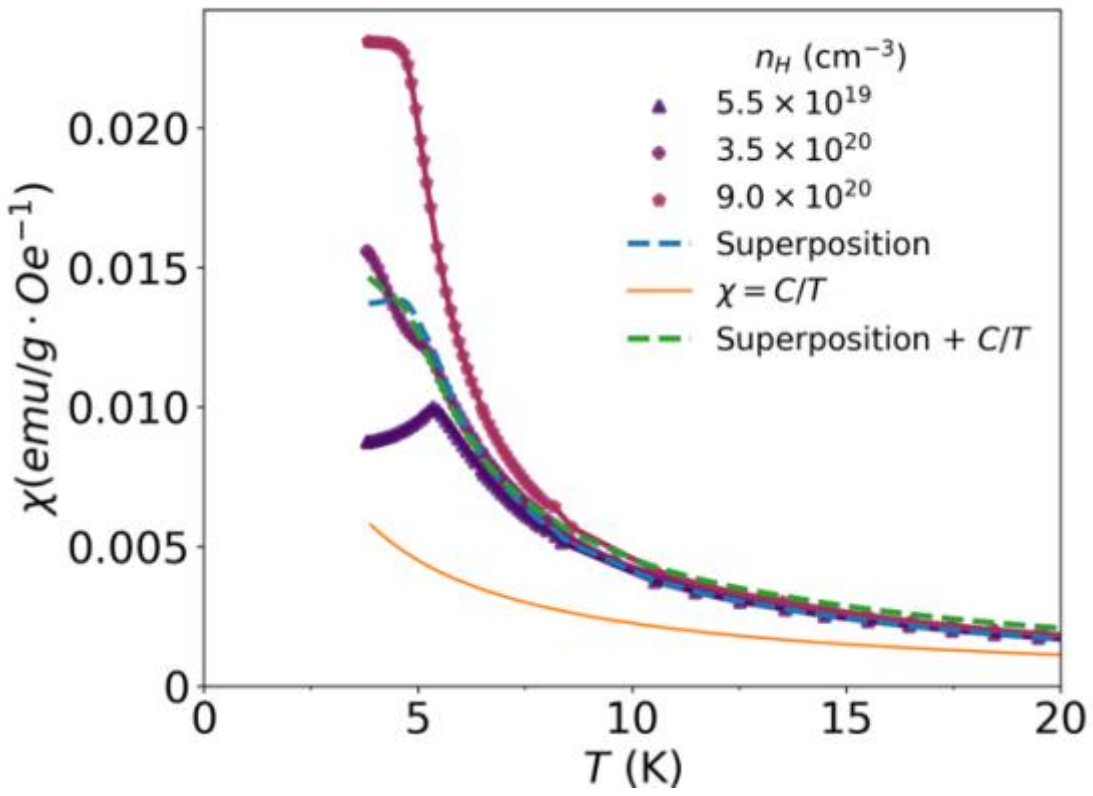

Figure S6. Magnetic susceptibility of three samples near the Hall concentration where the AFM-FM transition occurs. In order to test if the response of the sample with $n_H$ = 3.5 x $10^{20}$ cm$^{-3}$ can be interpreted as a mixed state of states with the two measured adjacent doping levels, its susceptibility is modeled as a linear superposition of the susceptibilities of samples with 5.5 x $10^{19}$ cm$^{-3}$ and 9.0 x $10^{20}$ cm$^{-3}$, depicted as a dashed blue line (see main text). The superposition of AFM and FM contributions alone does not capture the monotonic increase of $\chi$ at very low temperatures. Adding a paramagnetic contribution $\chi = C/T$ (orange curve), leads to a better description in the low-temperature region (dashed line).

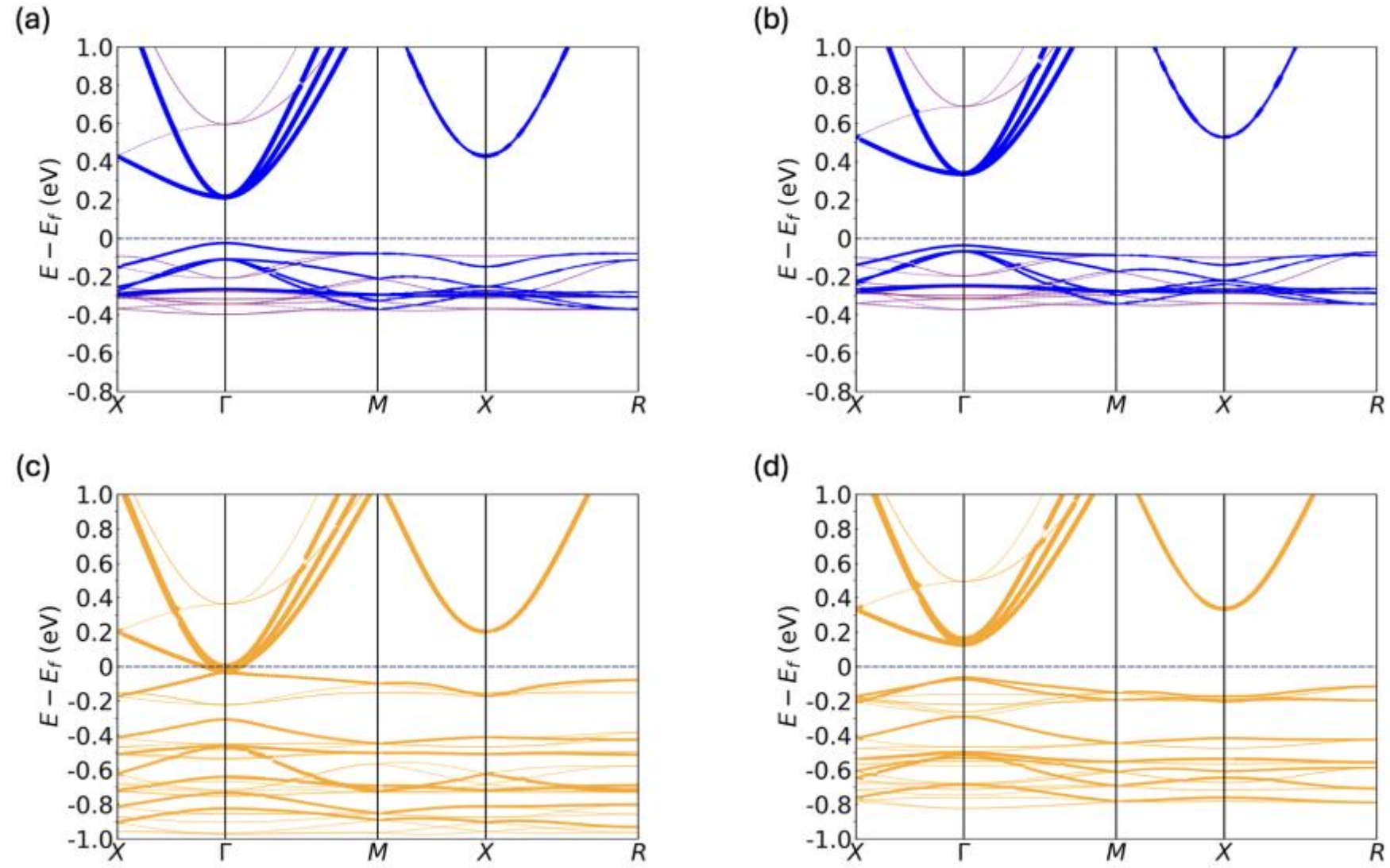

Figure S7. Unfolded band structure of undoped AFM ETO from DFT + $U$. (a) $U$ = 5.7 eV without SOC. (b) $U$ = 6.2 eV without SOC. (c) $U$ = 5.7 eV with SOC. (d) $U$ = 6.2 eV with SOC. The thin solid lines are folded band structures from the 2x2x2 supercell. The thick solid lines are unfolded band structures for a 5-atom unit cell. The energy zero is shifted to the Fermi energy $E_f$.

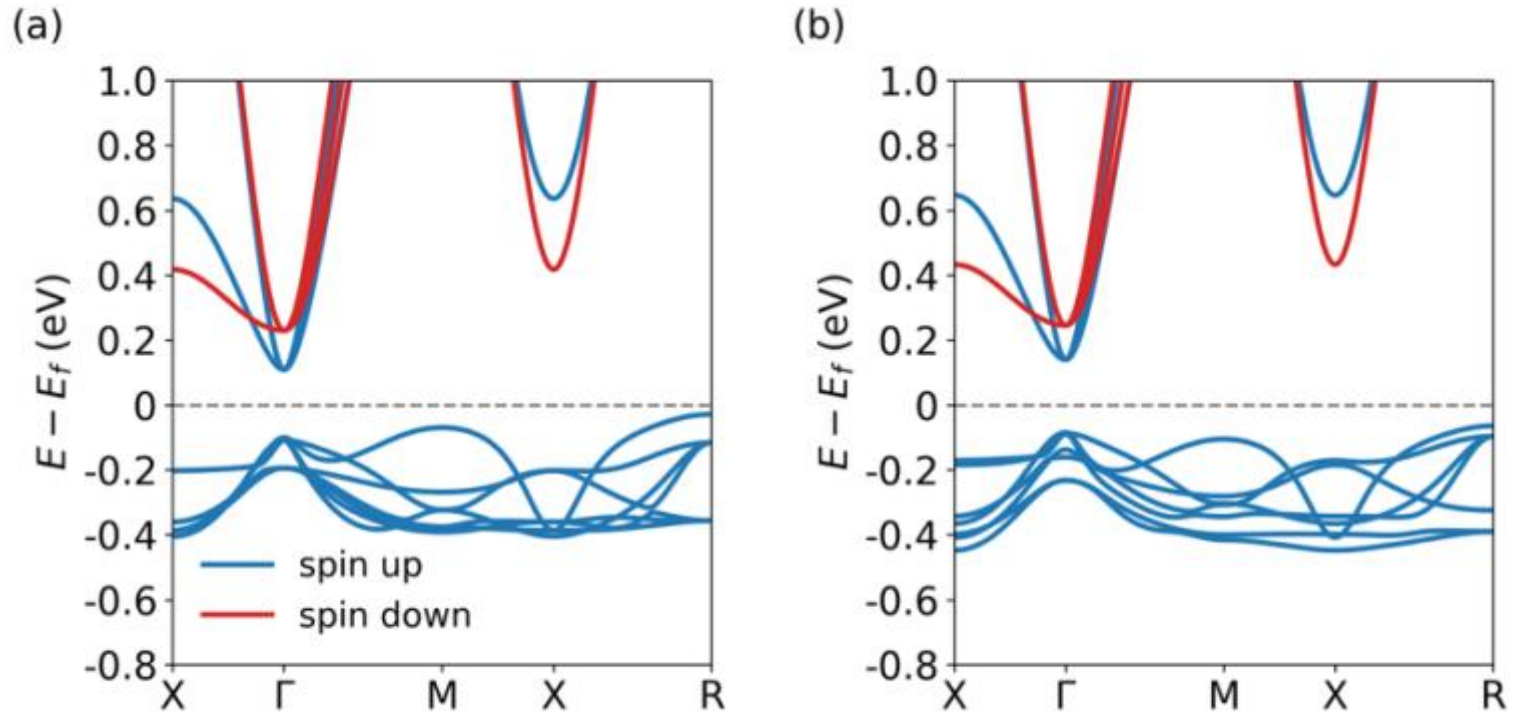

Figure S8. Band structure of undoped FM ETO without SOC from DFT + $U$. (a) $U$ = 5.7 eV. (b) $U$ = 6.2 eV. The energy zero is shifted to the Fermi energy $E_f$.

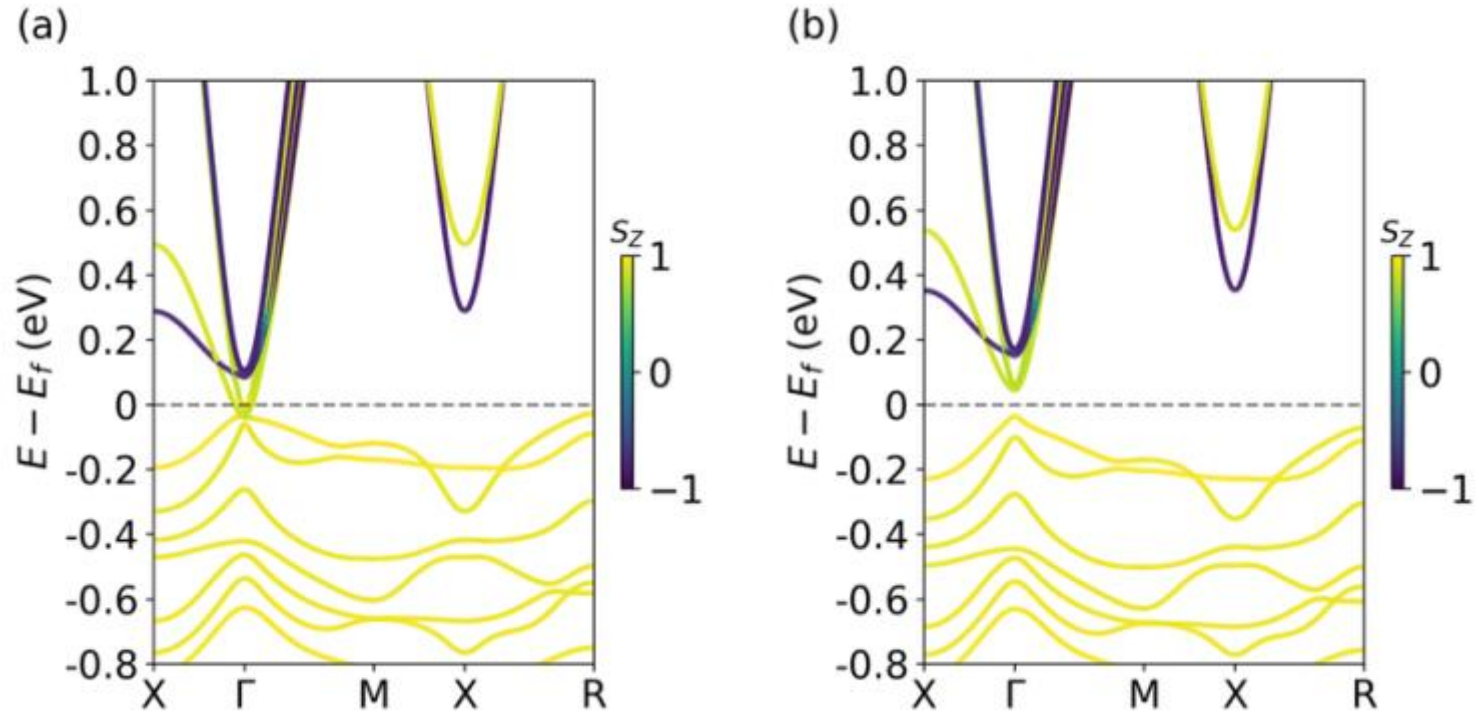

Figure S9. Spin-projected band structure of undoped FM ETO with SOC from DFT + $U$. (a) $U$ = 5.7 eV. (b) $U$ = 6.2 eV. The energy zero is shifted to the Fermi energy $E_f$.

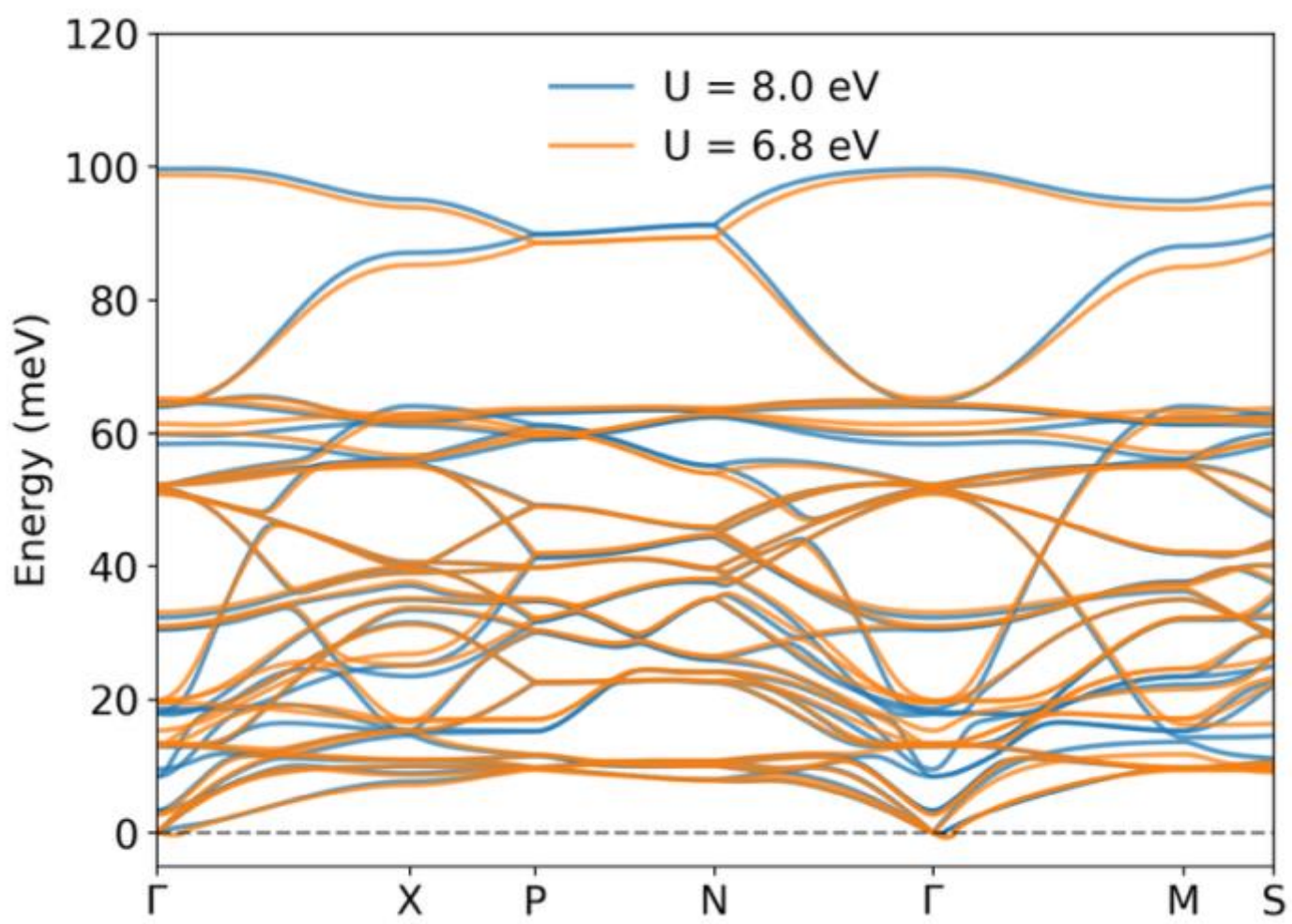

Figure S10. Phonon band structures of $EuTiO_3$, calculated from DFT with $U$ = 8 and 6.8 eV. The weak instability near the Gamma point is an error that emerge as a result of the manner in which the acoustic sum rule is treated and does not affect any of the results discussed in the present work.

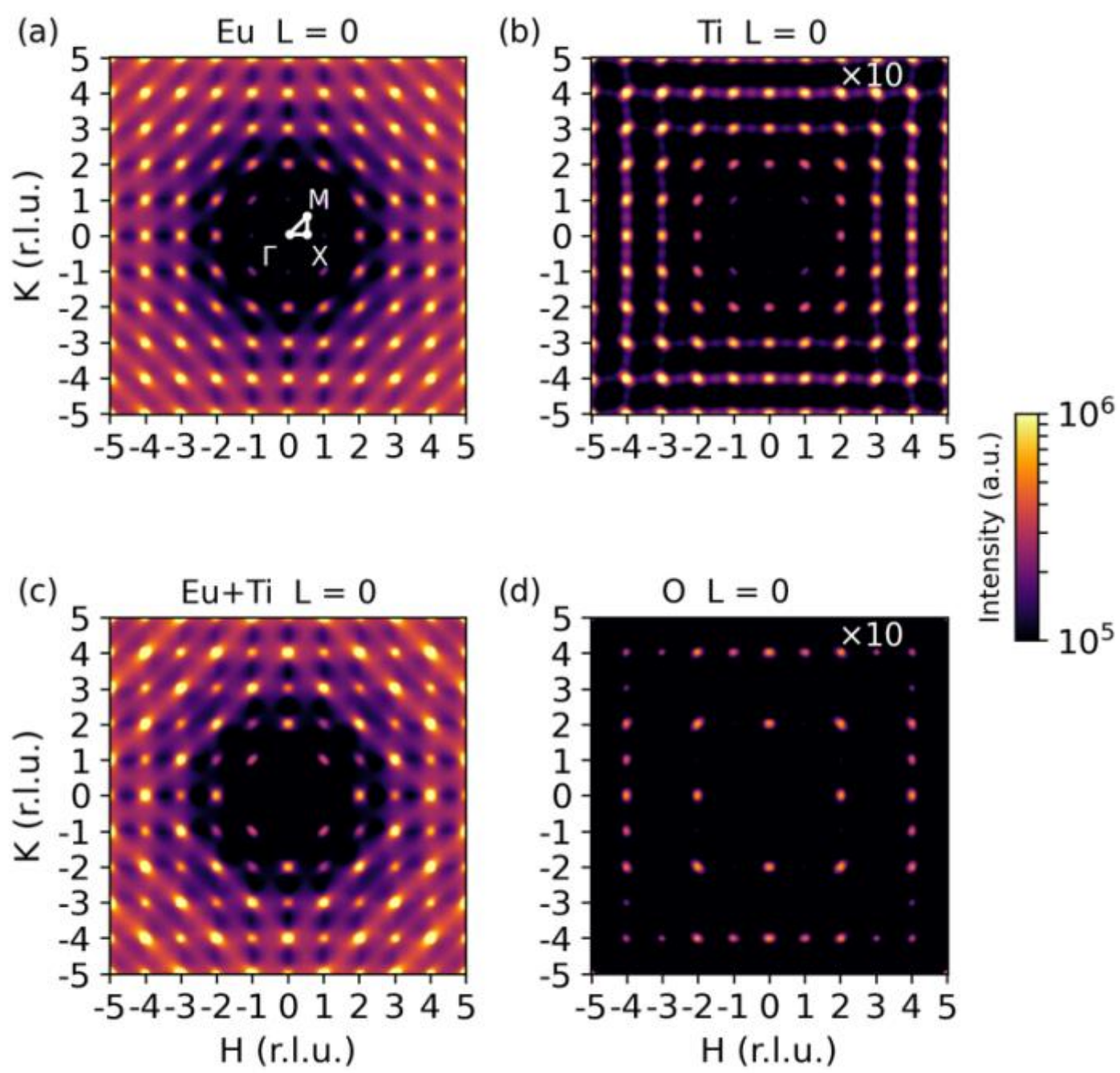

Figure S11. Thermal diffuse scattering simulation of decomposed elements in the *HK*0 plane. (a) Eu only; (b) Ti only; (c) Eu and Ti only; (d) O only. The decomposed thermal diffuse scattering patterns were calculated by assuming that the form factors of other elements are zero. For clarity, the Ti and O contributions in (b) and (d) are multiplied by a factor of 10. As shown, Ti and O contribute only at certain integer reciprocal lattice points based on the selection rules and do not contribute to thermal diffuse scattering streaks in the *HK*0 plane.

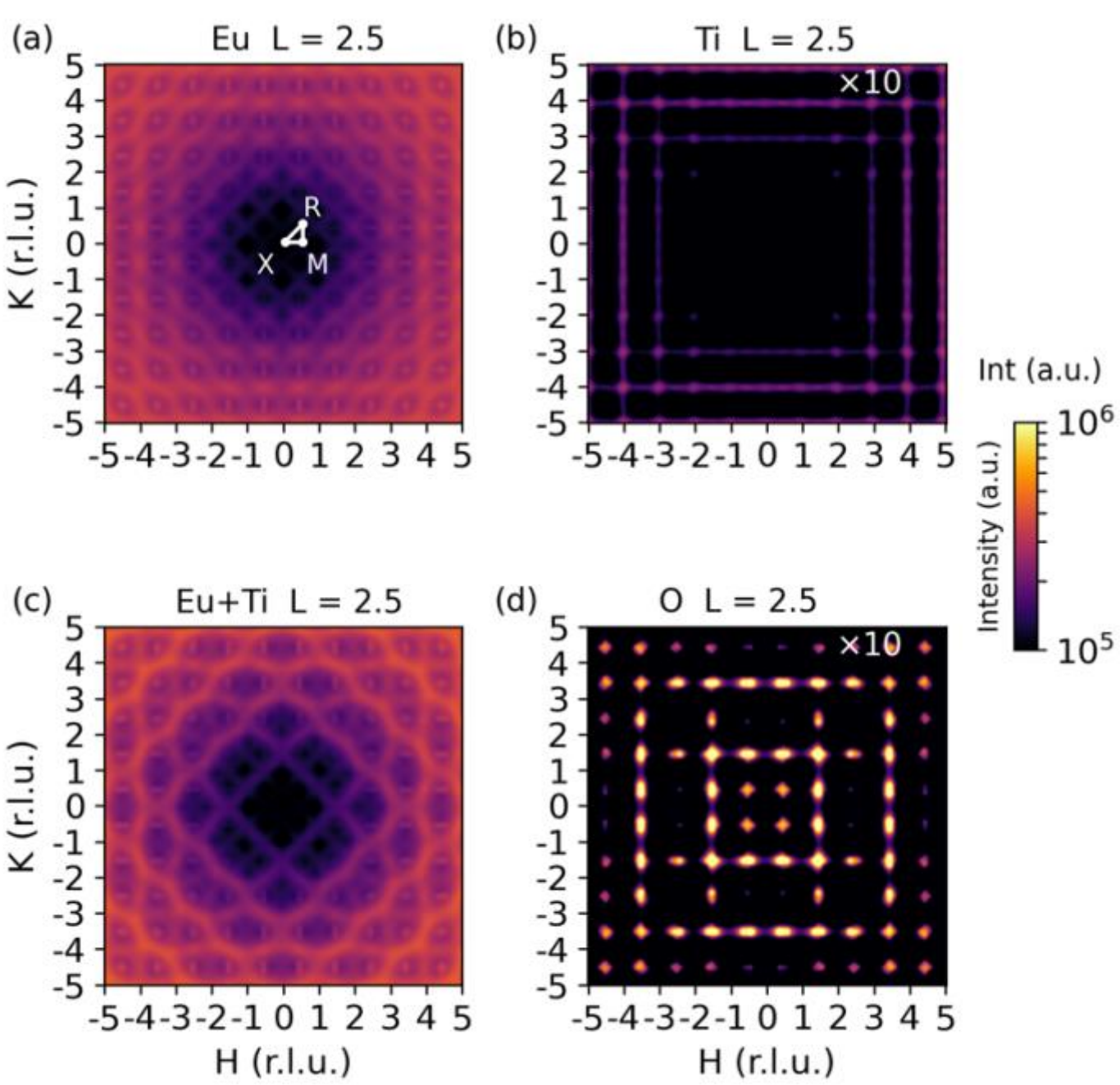

Figure S12. Thermal diffuse scattering simulation of decomposed elements in the *HK*2.5 plane. (a) Eu only; (b) Ti only; (c) Eu and Ti only; (d) O only. The decomposed thermal diffuse scattering patterns were calculated by assuming that the form factors of other elements are zero. For clarity, the Ti and O contributions in (b) and (d) are multiplied by a factor of 10. As shown, the combination of Eu and Ti results in the curvilinear pattern, and O contributes to the diffuse scattering at half-integer points in the *HK*2.5 plane.

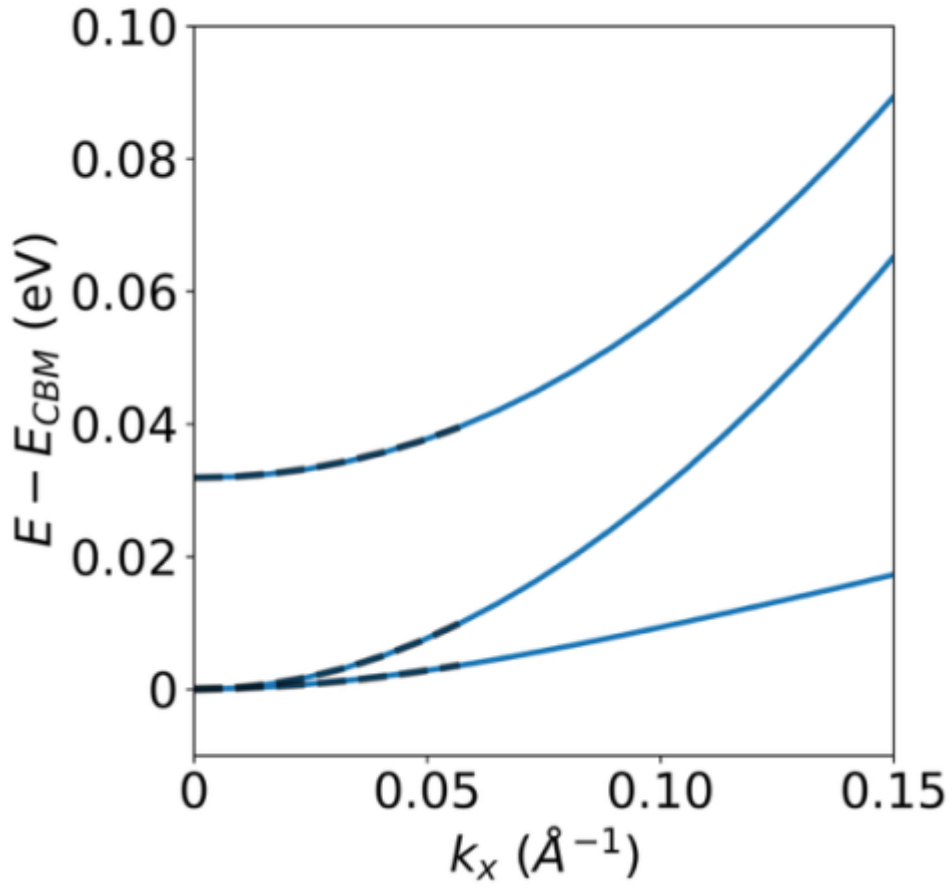

Figure S13. Fits to the lowest three conduction bands along the $\Gamma - X$ direction using the band structures of undoped AFM ETO using $U = 6.2$ eV with SOC. The blue solid lines are the DFT + $U$ band structure. The gray dashed lines are fitted parabolic curves. The fit range is from 0 to 0.06 $Å^{-1}$. The energy is shifted to the conduction band minimum (CBM).

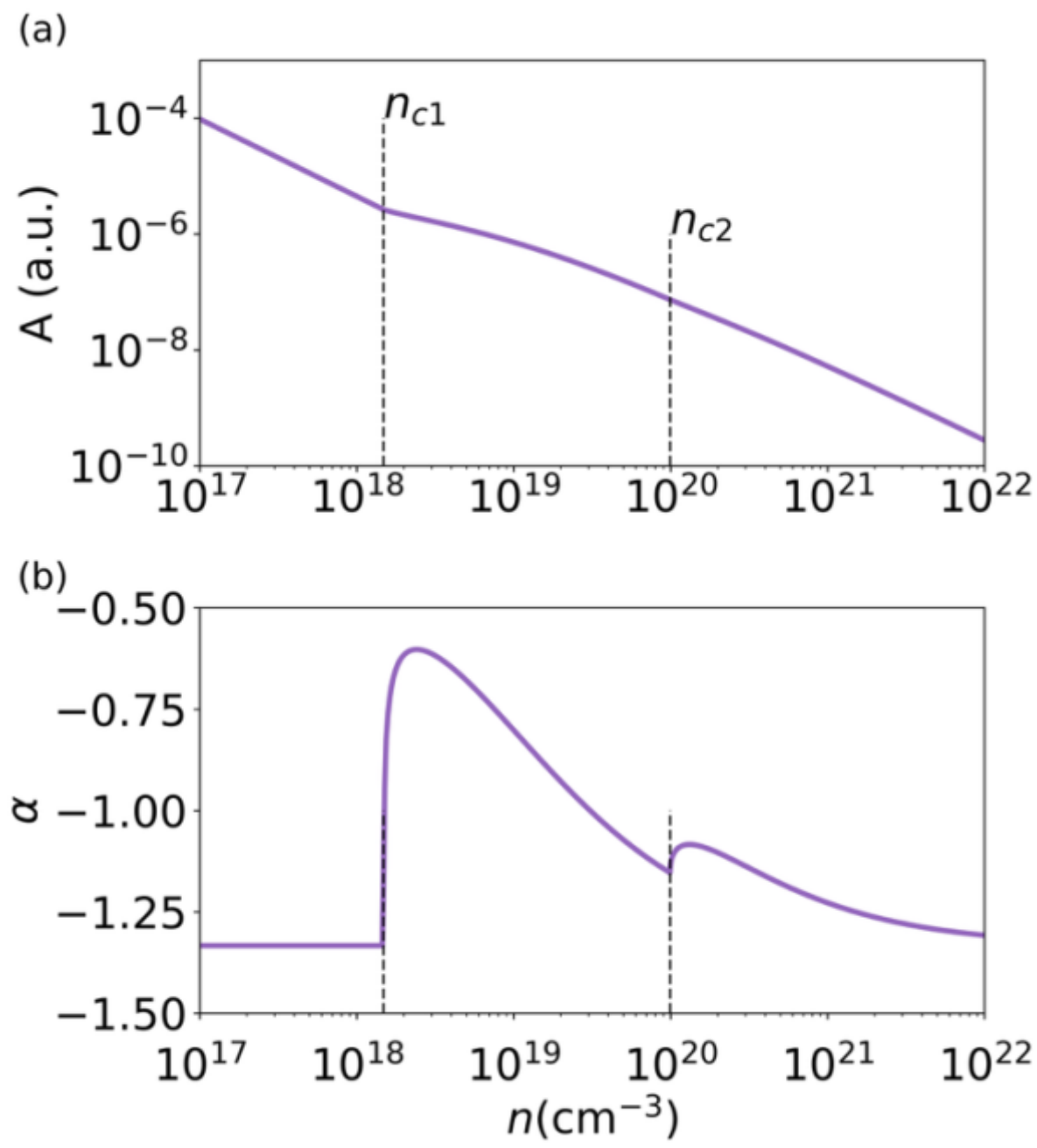

Figure S14. (a) $A(n)$ for the three-band model. (b) The corresponding exponent $\alpha$ in $A \propto n^{\alpha}$. The three-band model is used for the analysis in Fig. 4. The Lifshitz transition carrier concentrations are marked by vertical dashed lines.

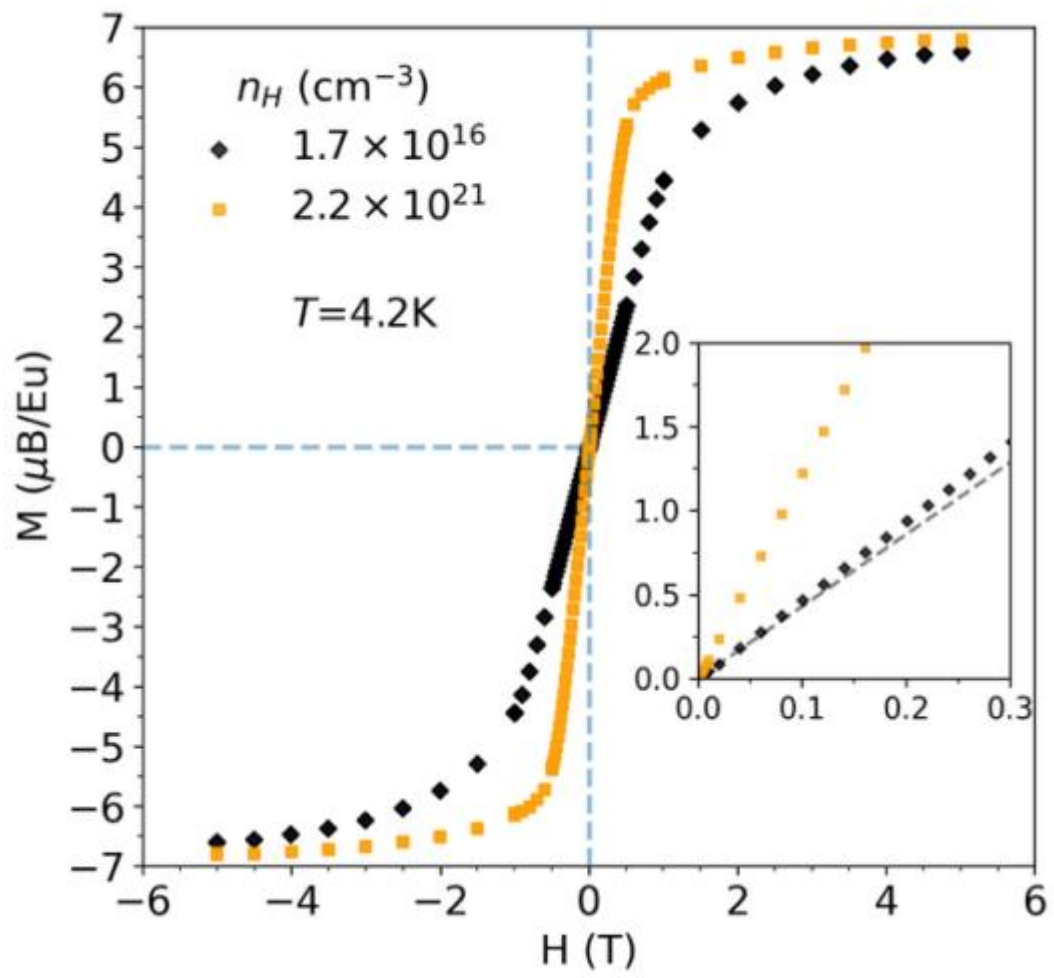

Figure S15. *M-H* data at 4.2 K for the as-grown sample and an OVD sample with the highest doping level. Inset: Zoom of low-field region. The gray dashed line is a guide to the eye, indicating its deviation from the linear behavior. For both samples, the magnetization gradually increases and approaches the ideal $Eu^{2+}$ ($S = 7/2$) spin moment of 7 μB at 5 T.

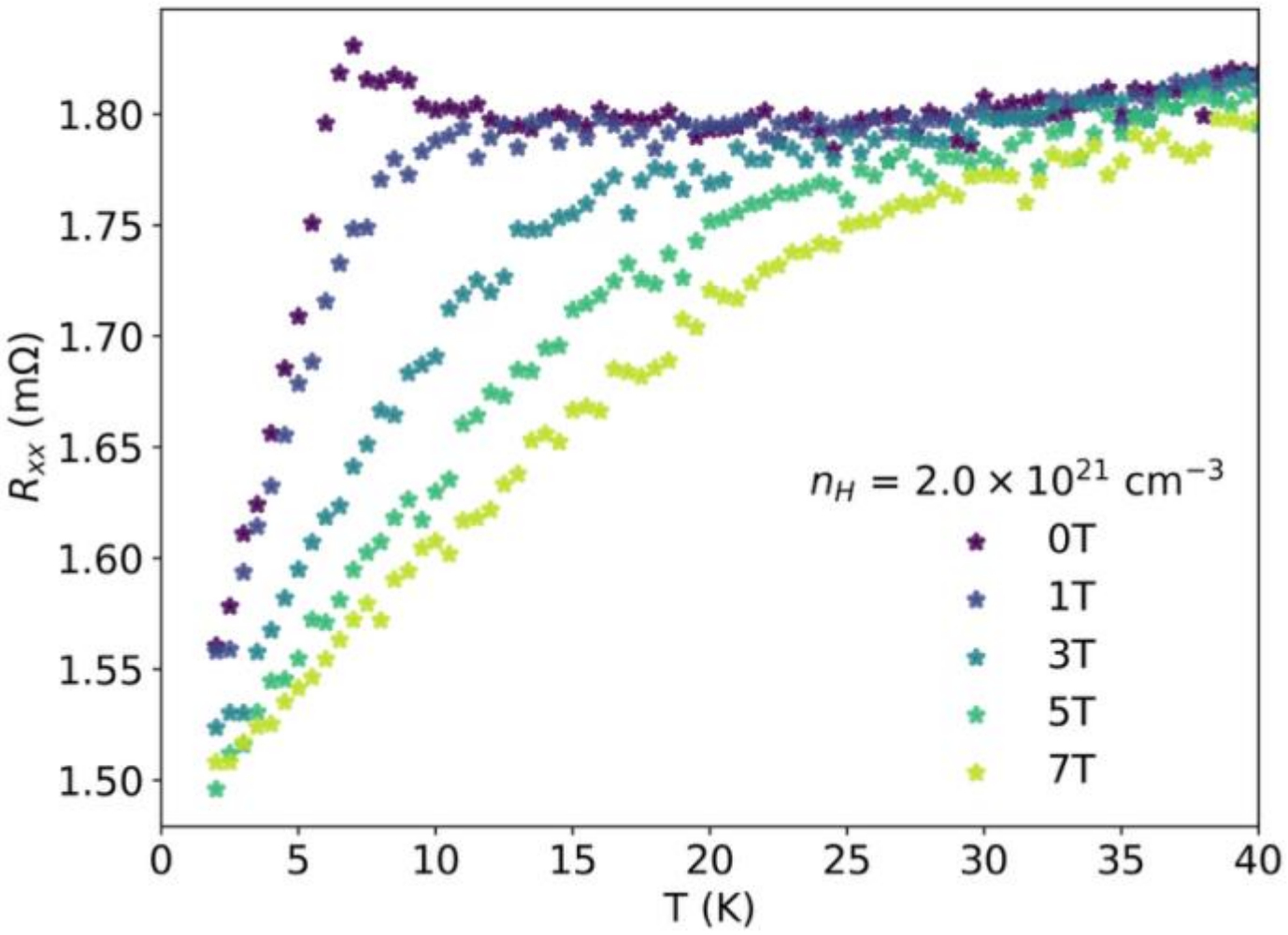

Figure S16. Longitudinal resistance of the $n_H$ = 2.0 x $10^{21}$ $cm^{-3}$ sample under different applied magnetic fields.

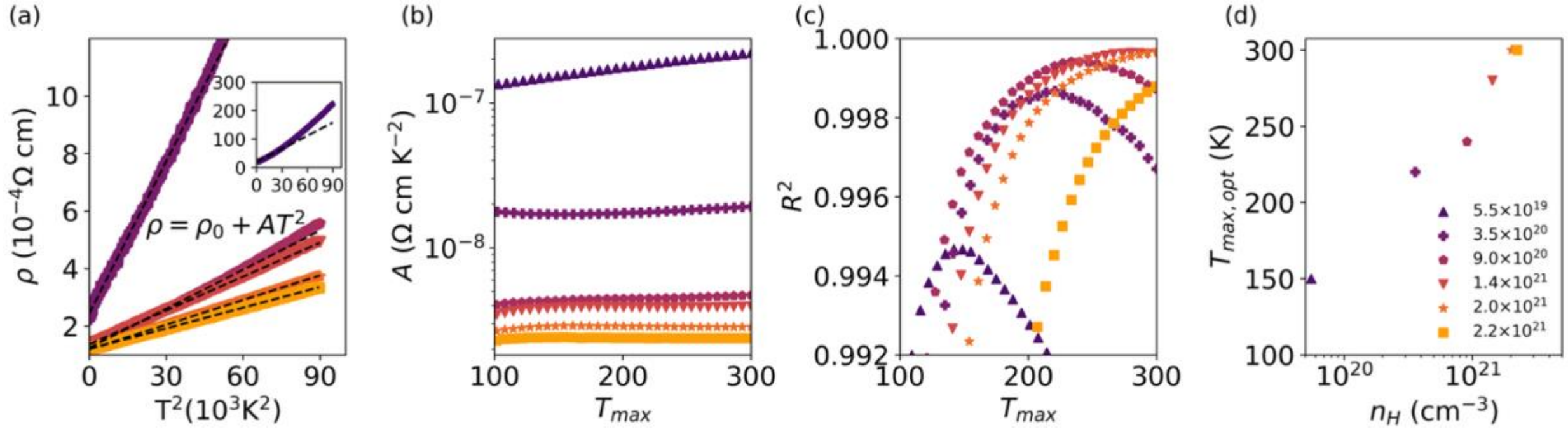

Figure S17. Resistivity fit results. The colors and the markers are the same as the ones in the main text Figs. 3, 4, and 5, and the carrier concentrations are shown in (d). (a) $\rho$ vs. $T^2$ and results of fits (dashed lines) to $\rho = \rho_0 + AT^2$ for the five highest doping levels of the present study. Inset: $\rho$ vs. $T^2$ and fit for the sixth highest doping level. (b) Fits were performed within a temperature range from 2 K to $T_{max}$. The fit parameter $A$ changes as the fit range (i.e., $T_{max}$) changes. (c) Goodness of fit. With increasing doping, the $R^2$ value is closest to 1 for $T_{max,opt}$ = 150, 220, 240, 280, 300, and 300 K. These "optimal" values were used in (a) and are shown in (d). Note that the maximum temperature of our measurements was 300 K, and thus this analysis is less reliable at the highest doping levels.

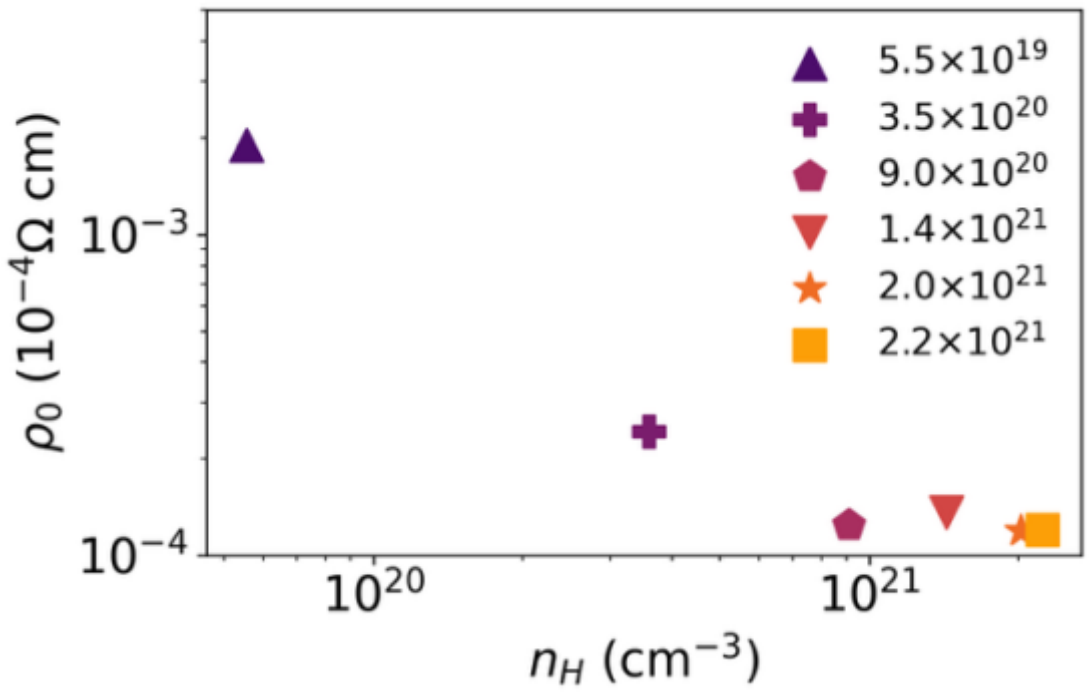

Figure S18. Fit results for the residual resistivity at the six highest doping levels. The colors and the markers are the same as the ones in main text Figs. 3, 4, and 5.